\documentclass[aps,prd,%twocolumn,
eqsecnum,showpacs,nofootinbib]{revtex4}

\usepackage{amsmath,amsfonts,amssymb,color,graphicx,latexsym,theorem,mathrsfs,comment
}

\newcommand{\ma}[1]{\mbox{$\mathcal{#1}$}}

\newcommand{\D}{{\rm d}}

\newcommand{\we}{\wedge}
\newcommand{\red}[1]{{\textcolor{red}{#1}}}

\definecolor{NoteColor}{rgb}{1,0,0}

\begin{document}

\title{Divergence equations and uniqueness theorem of static black holes}

%<<<<<<<<<<<<< AFFILIATION >>>>>>>>>>>>>>>%

\author{${}^{1}$Masato Nozawa, ${}^{2,3}$Tetsuya Shiromizu, ${}^{3,2}$Keisuke Izumi and ${}^{4}$Sumio Yamada}

\affiliation{${}^1$Yukawa Institute for Theoretical Physics, Kyoto University, Kyoto 606-8502, Japan}
\affiliation{${}^2$Department of Mathematics, Nagoya University, Nagoya 464-8602, Japan}
\affiliation{${}^3$Kobayashi-Maskawa Institute, Nagoya University, Nagoya 464-8602, Japan}
\affiliation{${}^4$Department of Mathematics, Gakushuin University, Tokyo 171-8588, Japan}

\begin{abstract}
Equations of divergence type in static spacetimes play a significant role in 
the proof of uniqueness theorems of black holes. We generalize the divergence 
equation originally discovered by Robinson in four dimensional vacuum spacetimes 
into several directions. We find that the deviation from spherical symmetry is 
encoded in a symmetric trace-free tensor $H_{ij}$ on a static timeslice. 
This tensor is the crux for the construction of the desired divergence equation, 
which allows us to conclude the uniqueness of the Schwarzschild black hole 
without using Smarr's integration mass formula. In Einstein-Maxwell(-dilaton) 
theory, we apply the maximal principle for a number of divergence equations to 
prove the uniqueness theorem of static black holes. In higher $(n\ge 5)$ dimensional 
vacuum spacetimes, a central obstruction for applicability of the current proof 
is the integration of the $(n-2)$-dimensional scalar curvature over the horizon 
cross-section, which has been evaluated to be a topological constant by the Gauss-Bonnet theorem for $n=4$. Nevertheless, it turns out that the $(n-1)$-dimensional  symmetric and traceless 
tensor $H_{ij}$ is still instrumental for the modification of the uniqueness proof 
based upon the positive mass theorem, as well as for the derivation of the 
Penrose-type inequality. 
\end{abstract}

\maketitle

\section{Introduction}

In four spacetime dimensions, asymptotically flat and static black holes to 
vacuum Einstein's equations are uniquely determined to be the Schwarzschild solution. 
A first proof was undertaken by Israel \cite{israel}, assuming that the horizon 
is spherical, non-degenerate and connected. The authors in \cite{MRS} were able 
to remove some technical conditions assumed in  \cite{israel} such as spherical 
topology, although this turned out to be a consequence of the topology theorem of event horizon.
Subsequently, Robinson \cite{robinson} gave a considerably simplified proof that 
encompasses the previous works. All of these methods are based upon nonlinear 
``divergence equations'' built out of the quantities on the static timeslice. 
Integrating this divergence equation over the static timeslice, one gets 
inequalities involving mass, area and surface gravity of the horizon and it turns 
out  that only the equalities are consistent. This leads to the spherical symmetry and 
therefore the metric is exhausted by the Schwarzschild solution. An alternative 
strategy proposed in \cite{bunting} makes an elegant use of the positive mass 
theorem \cite{Schon:1979rg,SchonYau,Witten:1981mf} and has been extended with 
suitable modifications into higher dimensions \cite{hwang,Gibbons:2002bh,Gibbons:2002av,Gibbons:2002ju,Rogatko:2003kj,Kunduri:2017htl}.

A notion closely linked to the black hole uniqueness is the Penrose 
inequality $M \ge \sqrt{A/(16\pi)}$ \cite{penrose1973},  where $A$ is the minimal 
area of the surface enclosing the apparent horizons and $M$ is the Arnowitt-Deser-Misner (ADM) 
mass of the asymptotically flat initial data set. Although the Penrose inequality 
is an important concept from the perspective of the cosmic censorship conjecture, 
its unequivocal proof is still lacking. Nevertheless, the Riemannian Penrose 
inequality has been established for an asymptotically flat three-manifold of 
nonnegative scalar curvature foliated by evolving surfaces of inverse mean curvature 
flow \cite{wald1977, imcf, jang1979}.\footnote{The conformal flow is another effective 
tool to prove the Riemannian Penrose inequality \cite{bray}. The proof has also been 
extended to the case with a charge in Ref. \cite{khuri2014}.} The monotonicity of 
the Geroch/Hawking quasi-local mass  \cite{hawking1968, geroch1973} along the 
inverse mean curvature flow is the key property for the proof of the Riemannian 
Penrose inequality. In this setting, the statement is also rigid, in the sense 
that equality is achieved if and only if the outside region of the apparent horizon 
is the Schwarzschild solution. This illustrates a fertile relationship between 
the uniqueness theorem of black holes and the Penrose inequality. Indeed, it has 
been argued that two inequalities in Israel's proof correspond to the Penrose-type 
inequality and its ``reversed'' version \cite{Mizuno:2009fj}, and the concurrent rigidities of the two inequalities correspond to the spherically symmetric spacetime. 

At the moment, the proof of the Riemannian Penrose inequality based on the inverse 
mean curvature flow fails in higher dimensions.  In order to understand in detail 
the relation to the Riemannian Penrose inequality, it is instructive to validate the uniqueness theorem in higher dimensions following the arguments 
in \cite{israel,MRS,robinson}. The proof would offer a new insight into the 
corresponding flow in higher dimensions and would be much more intuitive than the 
one  exploiting the positive mass theorem. However, it has been widely believed 
that the proofs in \cite{israel,MRS,robinson} do not have a simple generalization 
into higher dimensions, since the dimensionality of the spacetime enters the proof 
in the following manner. The proofs in \cite{israel,MRS,robinson} are based upon 
a divergence equation defined on a spacelike hypersurface which is asymptotically 
flat and terminates at the bifurcation surface of event horizon. This divergence 
equation gives rise to several inequalities involving integration of the 
scalar curvature for the induced metric on the horizon cross-section. In four 
dimensions, the Gauss-Bonnet theorem enables us to evaluate this quantity as a 
topological invariant. Obviously this is not possible in $n\ge 5$ dimensions. 
Moreover, the source term in the  divergence equation involves the Cotton tensor for the spatial metric,  which turns out to vanish in a static and  spherically symmetric spacetime. Since the Cotton tensor is an 
obstruction for the conformal flatness only for the three dimensional space, 
the existence of the desired divergence equation might be special to 
spatial dimension three.

These unsettled issues motivate us to study more deeply the uniqueness proofs based on the divergence equations. 
In hindsight, it is surprising that there exists a useful formula of divergence type adapted to the proof of the uniqueness theorem. We therefore attempt to provide a systematic derivation of the divergence equation toward the spherical symmetry. The present reformulation has several advantages over the original treatment in \cite{robinson}. Our formula in the vacuum case 
includes an additional free parameter, which allows us to conclude the uniqueness of the Schwarzschild solution 
without invoking the integrated mass formula that relates the mass, horizon area and surface gravity. 
The redundancy of the mass formula is a desirable presage when one tries to apply the proof for the non-asymptotically flat 
situation. For the Einstein-Maxwell theory, previous attempts for the uniqueness proof based upon divergence equation 
made a heavy use of the property of the symmetric coset space of the nonlinear sigma model. In contrast, our formulation does not rely on this property. 
In Einstein-Maxwell-dilaton theory, the most difficult problem was how to determine the value of the dilaton field at the horizon, which 
is not constant in general. We overcome this hindrance by discovering an entirely new divergence equation which is used to fix the dilaton field at the horizon in conjugation with the maximal principle. In addition, our formula in four dimensions does not make a direct appeal to the Cotton tensor to conclude the spherical symmetry. In place of the Cotton tensor, a central role is played by the symmetric and trace-free tensor $H_{ij}$ constructed out of geometric quantities on a spatial hypersurface. Although our proof does not admit a straightforward higher-dimensional generalization, this tensorial field is of considerable help in modifying the proof based upon the positive mass theorem. This presents a more geometrically intuitive explanation for the spherical symmetry, rather than the Dirichlet boundary value problem defined on the underlying Euclidean space. The Penrose-type inequality in the static case is also derived from the diverse divergence equations in higher dimensions.

The present article is organized  as follows. In the next section, we study the uniqueness 
proof of static vacuum black holes by extending the result of Robinson \cite{robinson}. 
In section \ref{sec:electrovac}, we shall discuss the proof in the electrovacuum case. 
The Einstein-Maxwell-dilaton theory will be addressed in section~\ref{sec:EMD}.
In section \ref{sec:higherD}, we will discuss the uniqueness theorem in higher dimensions.   
Concluding comments are described in the final section \ref{sec:summary}.

\section{Uniqueness theorem for vacuum black holes}
\label{sec:vacuum}

Let us consider the solutions to the vacuum Einstein equations $R_{\mu\nu}=0$. 
The most fundamental black hole solution is the Schwarzschild metric
\begin{align}
\label{Sch}
\D s^2=-\left(1-\frac{2M}{r}\right)\D t^2+\left(1-\frac{2M}{r}\right)^{-1}\D r^2+r^2 \D \Omega_2^2 \,,  
\end{align}
where $\D \Omega_2^2 =\D \theta^2+\sin^2\theta \D \varphi^2$ is the standard metric of a unit two-sphere. 
Here $M>0$ is the ADM mass \cite{Arnowitt:1962hi}. 
This metric is static, spherically symmetric and asymptotically flat. A regular event horizon locates at $r=2M$.

Among the static black hole solutions in the asymptotically flat spacetimes, the Schwarzschild solution is the 
only vacuum solution which admits a regular horizon~\cite{israel,MRS,robinson,bunting}. 
We discuss how the proof proceeds. 
When the spacetime admits a static Killing vector, the metric can be cast into the following form 
\begin{eqnarray}
\label{metric}
\D s^2=-V^2(x)\D t^2 +g_{ij}(x) \D x^i \D x^j \,, 
\end{eqnarray}
where $g_{ij}$ is the metric of the constant timeslice $\Sigma$. 
$V$ and $g_{ij}$ are independent of the time coordinate $t$. 
The event horizon locates at $V=0$, where the Killing vector $\partial/\partial t$ becomes null. The vacuum Einstein equations $R_{\mu\nu}=0$ decouple into
\begin{align}
\label{vacuumtt}
D^2V=0\,,
\end{align}
and 
\begin{align}
\label{vacuumij}
{}^{(3)}\!R_{ij}=\frac 1V D_iD_j V\,, 
\end{align}
where $D_i $ and ${}^{(3)}\!R_{ij}$  are the linear connection and the Ricci tensor 
associated with $g_{ij}$. 
Here and throughout this paper, we use the abbreviated notation $D^2 V=D_iD^iV$ and 
$(D V)^2=|D_iV|^2=D_iVD^iV$. 
From these equations, one sees that the scalar curvature for the space ($\Sigma, g_{ij}$)
vanishes.

For later purpose, it turns out useful to locally foliate $\Sigma$ by the level set $\ma S_V=\{V={\rm constant}\}$. 
Let us denote the unit normal to $\mathcal S_V $ in $\Sigma$ by 
\begin{align}
\label{}
n_i = \rho D_i V \,,  
\end{align}
where $\rho \equiv (D^iVD_iV)^{-1/2}$ stands for the lapse function. The induced metric and the extrinsic curvature of $\ma S_V$ in $\Sigma$ are given respectively by 
\begin{align}
\label{}
h_{ij}=g_{ij}-n_i n_j \,, \qquad k_{ij}=h_i{}^k D_k n_j \,.  
\end{align}
We shall denote  (twice) the mean curvature and the shear tensor of   $\ma S_V$  as 
\begin{align}
\label{}
k\equiv h^{ij}k_{ij} \,, \qquad \sigma_{ij}\equiv k_{ij}-\frac 12 k h_{ij} \,. 
\end{align}
The vacuum Einstein's equations can be  expressed in terms of these geometric objects. 
The equations which we need in our analysis are 
\begin{align}
\label{vacEineq_rhoR}
 n^i D_i \rho= \rho k \,, \qquad 
 {}^{(2)}\! R=\frac{2k}{V\rho}+k^2 -k_{ij}k^{ij} \,.  
\end{align}
Here $ {}^{(2)}\! R$ is the scalar curvature for the first fundamental form $h_{ij}$. 
The former equation stems from (\ref{vacuumtt}), while the latter is derived from (\ref{vacuumij}) and the Gauss equation.

Let us now specify our boundary conditions in terms of these geometric quantities. 
The curvature invariant $\ma K=R_{\mu\nu\rho\sigma}R^{\mu\nu\rho\sigma}$ 
is easily computed as 
\begin{align}
\label{}
\ma K= \frac{8}{V^2 \rho^2}\left[
k_{ij}k^{ij}+k^2+\frac{2}{\rho^2}(\mathcal D\rho)^2 
\right]\,,
\end{align}
where  $\mathcal D_i$ is the linear connection for $h_{ij}$. 
The finiteness of $\mathcal K$ at the horizon $V=0$ imposes the boundary conditions
\begin{align}
\label{vac_BChorizon}
\left.k_{ij}\right|_{V=0}=0 \,, \qquad \left. \mathcal D_i \rho \right|_{V=0}=0\,. 
\end{align}
The second condition represents the zero-th law, i.e, equilibrium condition, of black hole thermodynamics, 
since $\rho_{0}\equiv \rho|_{V=0}$ corresponds to the inverse of surface gravity of the event horizon. In what follows, we assume that the event horizon is nonextremal $(0<\rho_0<\infty)$ and connected.

The boundary conditions at infinity are fixed by the 
asymptotic flatness  
\begin{align}
\label{BCinf}
V \sim 1-\frac{M}{r} \,, \qquad g_{ij} \sim \left(1+\frac{2M}{r}\right)\delta_{ij} \,,
\end{align}
where $M(>0)$ is the ADM mass. 
In terms of geometric quantities of $\ma S_V$, the asymptotic boundary conditions 
(\ref{BCinf}) translate into
\begin{align}
\label{BCinfkrho}
k\sim \frac{2}{r} \,, \qquad \rho \sim \frac{r^2}{M} \,, \qquad \sigma_{ij}=O(1/r^6) \,. 
\end{align}
By the maximum/minimum principle, (\ref{vacuumtt}) fixes the range of $V$ as
\begin{align}
\label{Vrange}
0\le V <1.
\end{align}
In the original work of Israel~\cite{israel}, the global foliation of $\Sigma=\mathbb R\times \mathcal S_V$ has been postulated.
Henceforth, the topology of the cross section at infinity $\mathcal S_{V=1}$ and at the horizon 
$\mathcal S_{V=0}$ must be topologically homeomorphic, i.e, they are both $S^2$.  
In the present formulation, by contrast, we shall use equations (\ref{vacEineq_rhoR}), (\ref{vac_BChorizon}), (\ref{BCinfkrho}) only at the 
evaluation of surface integrals either at infinity or at the horizon. Thus, the local existence of the foliation at each neighborhood is sufficient for our purpose. 
This is a main advantage of the current prescription. 

\subsection{Robinson's proof}

Let us first recapitulate the argument in Robinson's short letter \cite{robinson}, 
where it was pointed out without derivation that there exists a powerful identity 
\begin{align}
\label{Robinsonid}
D_i \left[
-2 f_1^R(V)\frac{D^i \rho}{\rho^3 }+\frac{f_2^R(V)}{\rho^2} D^i V
\right]=\frac 14 \rho ^2 f_1^R(V)V^4 C_{ijk}C^{ijk} 
+\frac{3f_1^R(V)}{\rho^2}\left| \frac{D_i\rho}{\rho}-\frac{4V}{1-V^2}D_i V\right|^2\,,
\end{align}
where $f_{1,2}^R(V)$ are given by
\begin{align}
\label{}
f_1^{R}(V)=\frac{c_1V^2+c_2}{V(1-V^2)^3}\,, \qquad f_2^{R}(V)=-\frac{2c_1}{(1-V^2)^3} +\frac{6(c_1V^2+c_2)}{(1-V^2)^4}\,. 
\end{align}
Here $C_{ijk}=2 D_{[i}({}^{(3)}R_{j]k}-({}^{(3)}R/4)g_{j]k})$ is the Cotton tensor, whose 
vanishing is equivalent to the conformal flatness of the three surface $\Sigma$. 
$c_1$ and $c_2$ are arbitrary constants.
To ensure $f_1^{R}(V)\ge 0$, the constants $c_1$ and $c_2$ are chosen to take values in 
\begin{align}
\label{Robinson_c1c2}
c_1 +c_2 \ge 0 \,, \qquad c_2 \ge 0 \,. 
\end{align}
For this range of parameters, the right hand side of (\ref{Robinsonid}) is nonnegative. Using the Stokes theorem, one can transform 
the volume integral of (\ref{Robinsonid}) over the spatial slice $\Sigma$ into surface integral. 
The surface integral can be estimated by using (\ref{vacEineq_rhoR}), (\ref{vac_BChorizon}), 
(\ref{BCinf}), giving rise to
\begin{align}
\label{}
0\le -\frac{\pi}{2M}(c_1+c_2)- \left[-4\pi \chi c_2 \rho_0^{-1}+(6c_2-2c_1)\rho_0^{-3}A_H\right] \,,
\end{align}
where $A_H$, $\chi $ are respectively the area and the Euler number for the cross-section $B$ of the 
horizon 
\begin{align}
\label{Euler}
A_H= \int _B \D S \,, \qquad 
\chi = \frac 1{4\pi} \int_B  {}^{(2)} R \D S \,. 
\end{align} 

One sees that the choice $c_1=1$ and  $c_2=0$ satisfies 
(\ref{Robinson_c1c2}), for which one obtains the relation
\begin{align}
\label{Robinsonint2}
A_H \ge \frac{\pi}{4M} \rho^3_0\,.   
\end{align}
For the choice $c_1=-1$, $c_2=1$, one gets 
\begin{align}
\label{Robinsonint1p}
A_H \le \frac 12 \pi \rho_0^2 \chi \,. 
\end{align}
This is obviously compatible only for $\chi >0$. Since 
the compact, orientable and connected two-surface of positive Euler number is 
inevitably a sphere, one obtains $\chi=2$. This is consistent with the general results on the topology of stationary black holes \cite{Hawking:1971vc,Chrusciel:1994tr}. Thus (\ref{Robinsonint1p}) is reduced to 
\begin{align}
\label{Robinsonint1}
  A_H \le \rho_0^2 \pi \,.  
\end{align}
We have obtained two inequalities (\ref{Robinsonint2}), (\ref{Robinsonint1}), but 
we need one more relation between physical/geometrical parameters $(M, \rho_0, A_H)$. This can be obtained by integrating $D^2 V=0$ over $\Sigma$, yielding Smarr's integrated mass formula \cite{Smarr:1972kt} 
\begin{align}
\label{Smarr4Dvac}
\rho_0 M=\frac {1}{4\pi}A_H \,.
\end{align}
This relation arises from the integrability of the Killing equation 
($\nabla^\nu\nabla_\nu\xi^\mu=-R^\mu{}_\nu\xi^\nu$) for the static Killing vector $\xi=\partial/\partial t$.

Eliminating $\rho_0$ by use of (\ref{Smarr4Dvac}), 
(\ref{Robinsonint2}) yields the ordinary Penrose inequality
\begin{align}
\label{}
A_H \le 16 \pi M^2\,,
\end{align}
while (\ref{Robinsonint1}) gives the reversed inequality 
\begin{align}
\label{}
A_H \ge 16 \pi M^2 \,. 
\end{align} 
Compatibility demands that the equality must hold. One thus deduces from (\ref{Robinsonid}) to conclude 
 that 
\begin{align}
\label{vacCDrho}
C_{ijk}=0 \,, \qquad  \frac{D_i\rho}{\rho}-\frac{4V}{1-V^2}D_i V=0 \,. 
\end{align}
By the Lindblom identity \cite{Lindblom}
\begin{align}
\label{Csq}
C_{ijk}C^{ijk}=\frac{8}{V^4 \rho^4}\left(
\sigma_{ij}\sigma^{ij}+\frac{(\ma D \rho)^2}{2\rho^2} 
\right) \,, 
\end{align}
it turns out that the foliation $\ma S_V=\{V={\rm constant}\}$ is shear-free in $\Sigma$ and 
the second equation in (\ref{vacCDrho}) is solved to give 
$\rho(V) =4M/(1-V^2)^2$, where the integration constant has been settled by the asymptotic 
flatness. Equation (\ref{vacEineq_rhoR}) then implies that the scalar curvature for the two-dimensional metric $\hat h_{ij}=(M \rho)^{-1} h_{ij}$  is a positive constant, implying that $\hat h_{ij}$ is the
standard metric of a unit sphere. It turns out that the spacetime is spherically symmetric. A change of variable $r=2M/(1-V^2)$ casts the metric into the Schwarzschild solution (\ref{Sch}). 
 This completes the proof. 

\bigskip 
In contrast to Israel's original method~\cite{israel}, 
Robinson's proof does not assume the foliation $\Sigma=\mathbb R\times \ma S_V$ throughout the domain of outer communications.   
Accordingly, one does not a priori put any restrictions to the topology of black holes. 
If the horizon is not spherical, the foliation $ \ma S_V=\{V={\rm constant}\}$ obviously fails to cover the whole of domain of outer communications. 
This would be a nice property when one tries to generalize the proof in higher dimensions, since the possibility for the topology of higher dimensional black holes are much richer.

\subsection{Generalization of divergence equation}

In Robinson's proof, the divergence equation (\ref{Robinsonid}) plays a central role. 
Its effectiveness to deduce the black hole uniqueness without 
expending considerable effort is an appealing feature. At the same time, it remains enigmatic why this kind of desirable equations exists at all.
Also, it seems unlikely that one can generalize the proof in the presence of electromagnetic field in this original formulation, without specifying the functional relationship between the norm of the Killing vector and the electrostatic potential. 
The presence of the Cotton tensor within the formula is not convenient as well, if one tries to generalize the 
present scheme to higher dimensions. Finally, we would like to 
relinquish the Smarr relation since analogous integrated mass formulas do not exist
in the asymptotically (A)dS case.

Motivated by these issues, let us try to generalize the equation (\ref{Robinsonid}).
Inspecting (\ref{Robinsonid}), we wish to find  a 
current $J^i$ satisfying the divergence type equation
\begin{align}
\label{currentcons}
D_i J^i =(\textrm{terms of a definite sign}) \,. 
\end{align}
The right hand side of this equation consists of a sum of the tensorial norms which vanish 
for the Schwarzschild solution. We find that the candidates of this kind are the  symmetric tensor of the form
\begin{align}
\label{vac_Hij}
H_{ij} \equiv D_iD_j V-\frac{2V}{\rho^2 (1-V^2)}g_{ij}+\frac{6V}{1-V^2}D_i VD_j V\,, 
\end{align}
and the vector field 
\begin{align}
\label{vacHi}
H_i \equiv \frac{D_i \rho}{\rho}-\frac{4V}{1-V^2}D_i V \,. 
\end{align}
These spatial tensors satisfy
\begin{align}
\label{}
H_{ij}D^j V=-\rho^{-2}H_i \,, \qquad H^i{}_i=0 \,. 
\end{align} 
The vanishing of the tensor $H_{ij}$ for the Schwarzschild solution can be easily 
deduced by decomposing (\ref{vac_Hij}) into geometric quantities of $\ma S_V$ as
\begin{align}
\label{vac_Hij_dec}
H_{ij}=\rho^{-1}\sigma_{ij}-2\rho^{-2}n_{(i}\ma D_{j)}\rho +\frac{1}{2\rho}\left(k-\frac{4V}{\rho(1-V^2)}\right)(h_{ij}-2n_in_j)\,. 
\end{align}
It follows that the tensor $H_{ij}$ encodes the deviation from spherical symmetry. (See 
the discussion after (\ref{Csq}),  and also (\ref{Sch_Uni_Hmndechigh}) for $n\ge 5$ dimensional case. We also refer to (\ref{Dizeta}) and (\ref{RijHij}) for further geometric meaning of $H_{ij}$.)
This can be also seen by computing the Cotton tensor, which is now expressed as
\begin{align}
\label{Cottonvac}
C_{ijk}=\frac{2}{V^2}(2H_{k[i}D_{j]}V+\rho^{-2} H_{[i}g_{j]k}) \,.   
\end{align}
Therefore our current aim is to show $H_{ij}=0$ under the present boundary conditions.

As the first step, let us make an ansatz  for the current $J^i$ to be the following form
\begin{align}
\label{ansatz}
J^i =f_1(V)g_1(\rho) D^i\rho +f_2(V)g_2(\rho) D^iV \,,
\end{align}
where $f_{1,2}(V)$ and $g_{1,2}(\rho)$ are functions of each argument which will be fixed below. 
This separable form of the current is the same as (\ref{Robinsonid}). 
The divergence of this current is computed as
\begin{align}
\label{DJvac1}
D_iJ^i= (f_1'g_1+f_2 g_2')D^i \rho D_i V +f_1 g_1' (D \rho)^2+f_1g_1D^2\rho +f_2'g_2 (DV)^2\,,
\end{align}
where the prime denotes the differentiation with respect to the  single variable of the corresponding function.
Using 
\begin{align}
\label{laprho}
D^2 \rho &=-\rho^3 |D_iD_j V|^2 +\frac 1V D^iVD_i \rho +\frac 3\rho (D \rho)^2 \,, 
\end{align}
equation (\ref{DJvac1}) is rewritten into
\begin{align}
\label{}
D_iJ^i=& f_1(V)\rho^3 g_1(\rho )\left[
-|H_{ij}|^2+\frac{|H_i|^2}{\rho^2}\left(3+\frac{\rho g_1'(\rho)}{g_1(\rho)}\right)\right]
+H_iD^i V S_1+S_2 \,,
\label{vac_divJ_3dim}
\end{align}
where 
\begin{align}
\label{S1}
S_1=&\frac{\rho g_1(\rho) V f_1(V)}{1-V^2}\left[
\frac{1-V^2}{V}\left(\frac 1V+\frac{f_1'(V)}{f_1(V)}\right) +12 +\frac{8\rho g_1'(\rho)}{g_1(\rho)} 
+\frac{1-V^2}{V}\frac{f_2(V)}{f_1(V)}\frac{g_2'(\rho)}{g_1(\rho)} 
\right] \,, \\ 
\label{S2}
S_2=& \frac{4V}{(1-V^2)\rho^2}S_1+\frac{V^2 f_1(V)g_2(\rho)}{(1-V^2)^2 \rho^2 }
\left[\frac{(1-V^2)^2 f_2'(V)}{V^2 f_1(V)}-\frac{8\rho g_1(\rho)}{g_2(\rho)}
\left(3+\frac{2\rho g_1'(\rho)}{g_1(\rho)}\right)
\right] \,. 
%& \frac{4V}{(1-V^2)\rho}[g_1(\rho)f_1'(V)+f_2(V)g_2'(\rho)]+\frac{g_2(\rho)}{\rho^2}f_2'(V)+\frac{4f_1(V)}{(1-V^2)^2}
%\left[4V^2 g_1'(\rho)+(1+5V^2)\frac{g_1(\rho)}{\rho}\right]\,. 
\end{align}
Now we would like to render the right-hand side of (\ref{vac_divJ_3dim}) to have a 
definite sign and to vanish for the Schwarzschild solution. Since we cannot control the signs of the last two terms 
in (\ref{vac_divJ_3dim}), we require $S_1=S_2=0$. 
Inferring from the last term of (\ref{S1}), one needs either (i) 
$g_2'(\rho)\propto g_1(\rho)$ or (ii) $f_2(V)\propto \frac{V}{1-V^2}f_1(V)$ 
to render the equations decoupled. 
In this paper we will focus on the former case,\footnote{
Even if we adopt the option (ii), the final conclusion is identical. We shall not attempt to follow this route.  
}
for which 
\begin{align}
\label{}
g_1(\rho)=-c\rho ^{-(c+1)} \,, \qquad g_2(\rho)= \rho ^{-c}\,,
\end{align}
where $c$ is an integration constant. 
Substituting these back into (\ref{S1}) and (\ref{S2}) yields two first-order linear differential equations 
\begin{align}
\label{}
f_1'(V)+f_2(V)+\frac{1+(3-8c)V^2}{V(1-V^2)}f_1(V)=0 \,, \qquad 
f_2'(V)+\frac{8c(1-2c)V^2 }{(1-V^2)^2}f_1(V)=0 \,.
\end{align}
These equations are easily solved to give 
\begin{align}
\label{}\
f_1(V)=\frac{(1-V^2)^{1-2c}}{V}\left[a+b(1-V^2)\right]\,, \qquad 
f_2(V)=\frac 2{(1-V^2)^{2c}}[a(2c-1)+2bc(1-V^2)] \,,
\end{align}
where $a$ and $b$ are integration constants. 
Using 
\begin{align}
\label{vac_Cottonsq}
\left|D_{[i}VH_{j]k}-\frac 1{2\rho^2} H_{[i}g_{j]k}\right|^2 
=\frac{1}{2\rho^2} \left[
|H_{ij}|^2-\frac{3}{2\rho^2}|H_i|^2
\right]\,, 
\end{align}
we finally arrive at an improved divergence equation
\begin{align}
\label{vac_divid}
D_i J^i =\frac{cf_1(V)}{2\rho^c}\left[
\left|2\rho^2 D_{[i}VH_{j]k}-H_{[i}g_{j]k}\right|^2 
+\left(2c-1 \right)|H_i|^2 
\right] \,.
\end{align}
That this formula contains three tunable parameters is the key to the 
present proof for the uniqueness. By writing the first term in terms of the Cotton tensor (\ref{Cottonvac}),  
this recovers Robinson's equation (\ref{Robinsonid}) for $c=2$, 
and equations (2.12) and (2.13) in \cite{MRS} for $c=1/2$.  The 
right-hand side of the above equation becomes positive semi-definite, 
provided\footnote{
We also need $0<\rho<\infty$ in the interior of $\Sigma$ for the right-hand side 
of (\ref{vac_divid}) to be well-defined. This can be shown by applying the 
maximal principle to (\ref{laprho}), as demonstrated in \cite{hwang}. 
}
\begin{align}
\label{vac_cdond}
f_1(V) \ge 0\,, \qquad c\ge \frac 12 \,. 
\end{align}
The condition $f_1(V)\ge 0$ for $0\le V < 1$ is assured by 
\begin{align}
\label{vac_abcond}
a\ge 0 \,, \qquad a+b \ge 0 \,. 
\end{align}

By Stokes' theorem, 
integration of (\ref{vac_divid}) over $\Sigma$ is transformed into the inequality for surface integrals 
\begin{align}
\label{vac_surfaceint}
\int _{\Sigma}D_iJ^i \D \Sigma =
\int_{S^\infty} J_i n^i \D S-\int_{B} J_i n^i \D S \ge 0 \,,
\end{align}
where $S^\infty$ is the two-surface at infinity and $B$ denotes the bifurcation two-surface
of the horizon. Upon using (\ref{vacEineq_rhoR}), (\ref{vac_BChorizon}), (\ref{BCinf}), 
we end up with 
\begin{align}
\label{vac_ineq}
0\le a [A_H \rho_0^{-(1+c)}- \pi (4M)^{1-c}] + (a+b)c \rho_0^{-(1+c)} [\pi \chi \rho_0^2-2A_H] \,. 
\end{align}
This inequality holds for any values of $a$, $b$ and $c$ satisfying (\ref{vac_cdond}) and (\ref{vac_abcond}), 
if and only if the pair of inequalities
\begin{align}
\label{vac_ineq2}
\pi \left(\frac{4M}{\rho_0}\right)^{1-c} \le \frac{A_H} {\rho_0^2}  \le\frac{ \pi}{2} \chi  \, 
\end{align}
is satisfied.

The case for $c=1$ gives $\chi \ge 2$, implying that only the spherical topology ($\chi=2$) is allowed.  By setting $\chi=2$, one sees that  the only case for $A_H=\pi \rho_0^2$ is consistent with (\ref{vac_ineq2}) for $c=1$. 
Hence, the inequality (\ref{vac_surfaceint}) is converted to an equality, and then 
(\ref{vac_divid}) implies $H_{ij}=0$.  Therefore,  the spacetime admits spherical symmetry, 
as desired. 

\bigskip
We have shown the uniqueness of the Schwarzschild solution without using Smarr's formula (\ref{Smarr4Dvac}). 
The underlying reason behind this is that one can choose $c=1$ for which 
the ADM mass disappears from (\ref{vac_ineq2}). This would not have been possible in the original 
form of Robinson's equation (\ref{Robinsonid}), since it corresponds to $c=2$. 
Even though this freedom does not offer any advantages in the proof of Schwarzschild solution, the unnecessity of the Smarr formula is a desirable precursor to the proof of uniqueness theorem via divergence equation in the asymptotically (A)dS case. 

It is worth commenting that the inequality $\chi \ge 2$ obtained here is the stronger than 
the one derived by the variational formula \cite{Hawking:1971vc}.
In particular this rules out explicitly the topology of  $\mathbb{RP}^2$ surface which has $\chi=1$ but is unorientable.

We also dispensed with the Cotton tensor in the source term of our divergence equation (\ref{vac_divid}), 
since the obstruction for spherical symmetry is completely encoded in $H_{ij}$. 
This property continues to be valid in higher dimensions, as we will see in section~\ref{sec:higherD}. 

From the separable form (\ref{ansatz}) of the current $J^i$, the geometric meaning is less obvious and it is hard to read off the spherical symmetry. This is rectified 
by recasting the current $J^i$ into a more suggestive form
\begin{align}
\label{}
J^i=- [(1-V^2)^2\rho]^{-c} \left[\frac{c}{V}(1-V^2)[a+b(1-V^2)] H^i +2aD^i V \right]\,. 
\end{align}
It is important to observe that $H_i$ defined in (\ref{vacHi}) is given by the derivative of 
$(1-V^2)^2\rho$. For the Schwarzschild solution, this is indeed constant 
$(1-V^2)^2\rho=4M$.  Because of this fact along with the equations of motion $D^2V=0$, 
the conservation of the current for the spherical symmetry becomes compelling.

We also remark that $H_{ij}=0$ is equivalent to the condition that 
$\zeta_i\equiv (1-V^2)^{-3}D_iV$ is a conformal Killing vector on $\Sigma$:
\begin{align}
\label{Dizeta}
 D_i\zeta_j +D_j \zeta_i-\frac{2}{3}D_k\zeta^k g_{ij} =\frac{2}{(1-V^2)^3}H_{ij}\,. 
\end{align}
For $H_{ij}=0$, 
this conformal Killing vector field corresponds to the dilatation vector field, which is always present in the spherically symmetric space. Namely, the conformally flat space $\D s_3^2=\D r^2/f(r)+r^2 (\D \theta^2+\sin^2\theta \D \phi^2)$ admits a dilatational conformal Killing vector of the form  $\zeta_i =(r/\sqrt{f(r)})D_i r$. 
This provides another geometric meaning to $H_{ij}$ as an obstacle for the existence of the conformal Killing vector corresponding to the dilatation, in a space of nonnegative scalar curvature.

\section{Einstein-Maxwell theory}
\label{sec:electrovac}

This section discusses the Einstein-Maxwell system described by the Lagrangian
\begin{align}
\label{}
L= R -F_{\mu\nu}F^{\mu\nu} \,. 
\end{align}
Here $F_{\mu\nu}=F_{[\mu\nu]}$ is the Faraday tensor. 
Consider the  static spacetime (\ref{metric}) and assume that the Maxwell field is  invariant under the action generated by the static Killing vector $\mathcal L_{\partial/\partial t} F_{\mu\nu}=0$.\footnote{
In general, if the four-dimensional spacetime admits a non-null Killing vector $\xi$, the Maxwell field satisfies 
$\ma L_\xi F=\Psi \star F$, where 
$(\star F)_{\mu\nu}=\frac 12 \epsilon_{\mu\nu\rho\sigma}F^{\rho\sigma}$
and $\Psi $ being constant \cite{Michalski,Ray}. We refer the readers to \cite{Tod:2006mp} for an attempt to get rid of the assumption of symmetry inheritance $\ma L_\xi F=0$ in the uniqueness proof.} 
Performing the electromagnetic duality rotation if necessary, the Bianchi identity $\D F=0$ brings the Maxwell field to be electric
\begin{align}
\label{}
F= \D t \we \D \psi \,, 
\end{align}
where $\psi $ is an electrostatic potential and we can work in a gauge in which $\psi$ is $t$-independent.  Since we are focusing on the asymptotically flat spacetime satisfying the null convergence condition, the topological censorship holds and therefore the domain of outer communication is simply connected~\cite{Chrusciel:1994tr,Galloway1995}, for which the global existence of $\psi$ is assured. 
The electrovacuum Einstein's equations then read
\begin{align}
\label{}
D^2 V=\frac 1V(D \psi)^2 \,, 
\end{align}
and 
\begin{align}
\label{}
{}^{(3)}R_{ij}-\frac 1V D_iD_j V=\frac 2{V^2}\left(-D_i \psi D_j\psi+\frac 12 (D \psi)^2 g_{ij} \right)\,.
\end{align}
The Maxwell equation gives 
\begin{align}
\label{Maxwelleq}
D_i (V^{-1}D^i \psi )= 0\,. 
\end{align}
Expressed in terms of geometric quantities for the foliation $\Sigma\simeq \mathbb R\times \ma S_V$, 
the relevant  Einstein's equations are 
\begin{align}
n^iD_i \rho=&\rho k -\frac{\rho^2}{V}[(n^iD_i \psi)^2+(\ma D \psi)^2] \,, \qquad 
\label{EV_trGausseq}
{}^{(2)} R =- k_{ij}k^{ij}+k^2+\frac{2k}{V\rho }-\frac{2}{V^2 }(\mathcal D\psi)^2 
+\frac{2}{V^2} (n^i D_i \psi)^2 \,. 
\end{align}

At infinity, the metric and the gauge field 
behave as  
\begin{align}
\label{EV_BCinf}
V \sim 1-\frac{M}{r} \,, \qquad g_{ij} \sim \left(1+\frac{2M}{r}\right)\delta_{ij} \,, \qquad 
\psi \sim  \frac{Q}{r} \,,  
\end{align}
where $M$ is the ADM mass and $Q$ is the electric charge which is taken to be positive without loss of generality.  In terms of $k$ and $\rho$, the asymptotic conditions can be translated as (\ref{BCinfkrho}). 
We assume that these conserved charges strictly obey the Bogomol'ny inequality \cite{Gibbons:1982fy,Nozawa:2014zia} 
\begin{align}
\label{}
M>Q\,. 
\end{align}
The boundary conditions at the event horizon $V=0$ can be determined by 
requiring the curvature invariants $R_{\mu\nu\rho\sigma}R^{\mu\nu\rho\sigma}$ 
and $F_{\mu\nu}F^{\mu\nu}$ to remain finite, leading to  
\begin{align}
\label{EV_BCH}
\left. k_{ij}\right|_{V=0}=0\,,
 \qquad \left.\mathcal D_i \rho\right|_{V=0} =0 \,, \qquad 
 \left.n^i D_i\psi\right|_{V=0}=0\qquad 
 \left.\mathcal D_i \psi\right|_{V=0}=0 \,. 
\end{align}
The constancy of $\rho$ and $\psi$ over the horizon means the zeroth law 
of black hole thermodynamics. 
The nonextremality condition of the horizon amounts to 
$0<\rho_0=\rho|_{V=0}<\infty$.
%and $\psi_0=\psi|_{V=0}\ne 0$.
%\red{[KI: Should this be $\psi_0=\psi|_{V=0}<1$?]} 
An example of the black hole solutions is the Reissner-Nordstr\"om solution
\begin{align}
\label{RN}
\D s^2=-f(r)\D t^2+f^{-1}(r)\D r^2+r^2 \D \Omega_2^2\,, \qquad 
\psi = \frac{Q}{r}\,, 
\end{align}
where $f(r)=1-2M/r+Q^2/r^2$. The uniqueness of (\ref{RN}) was first addressed by Israel \cite{Israel:1967za}  under the assumption that the horizon is connected, nondegenerate and spherical. Later works \cite{MR,Simon1984,Ruback,MuA1992} removed some of these assumptions.  
We are now going to show the uniqueness of the (\ref{RN}), based on the divergence equation. 
A primary utility of this scheme is that divergence equations are suitable for adaptation of the 
maximum/minimum principle.

From the field equations, one can derive following two divergence equations 
\begin{align}
\label{diveq1}
D_i(D^i V -V^{-1}\psi D^i \psi)=&0 \,, \\
\label{diveq2}
D_i \left(\frac{V^2+\psi^2}{V}D^i \psi -2\psi D^i V\right)=&0 \,. 
\end{align}
Upon integrating (\ref{Maxwelleq}), (\ref{diveq1}) and (\ref{diveq2})
over the static timeslice $\Sigma$,  one easily obtains
\begin{align}
\label{}
 4\pi Q+\int _B \frac{n^i D_i \psi}{V} \D S =& 0 \,, \\
 4\pi M-\rho_0^{-1} A_H + \psi_0 \int_B  \frac{n^i D_i \psi}{V} \D S=& 0 \,, \\
 4\pi Q-2 \rho_0^{-1} \psi_0 A_H+\psi_0^2 \int_B  \frac{n^i D_i \psi}{V} \D S=& 0 \,. 
\end{align}
The above equations are combined to give Smarr's mass formula
\begin{align}
\label{Smarr:EV}
A_H =4\pi \rho_0 (M-Q\psi_0)\,, 
\end{align}
as well as
\begin{align}
\label{}
\psi_0=\frac{M-\sqrt{M^2-Q^2}}{Q} \,,
\end{align}
where the sign in front of the square root in $\psi_0$ has been chosen to ensure 
$A_H=4\pi \rho _0 \sqrt{M^2-Q^2}>0$.

Of crucial importance to the present proof is the following equation
\begin{align}
\label{G1eq}
D^2G_1-D_i [\log (V\psi^{-2})]D^i G_1=0 \,,
\end{align}
where $G_1(V,\psi) \equiv \psi^{-1}(1-V^2)+\psi $. Here note that (\ref{Maxwelleq}) and 
the maximum/minimum principle \cite{elliptic} tell us $0<\psi<\psi_0$ and then 
$|D_i \log (V\psi^{-2})|<\infty $. Hence, the maximum/minimum principle can be applied to (\ref{G1eq}), implying that  $G_1(V,\psi)$ never acquires the global maximum/minimum 
in the interior of $\Sigma$. From the present boundary conditions,  the values of $G_1(V,\psi) $ 
at the horizon and at infinity both coincide to give $2M/Q$, and hence $G_1(V,\psi)$ 
is constant ($=2M/Q$) throughout $\Sigma$. Then $\psi $ depends only on $V$ and is given by 
\begin{align}
\label{RN_psiV}
\psi =\frac{1-\beta(V)}{q} \,, 
\end{align}
where for notational simplicity we have introduced 
\begin{align}
\label{}
\beta(V) \equiv \sqrt{1-q^2(1-V^2)} \,, \qquad q\equiv Q/M<1 \,. 
\end{align}

One can also derive the relation (\ref{RN_psiV}) by several fashions. 
Israel exploited a divergence equation which depends explicitly on physical parameters 
$M$, $Q $ to conclude (\ref{RN_psiV}) \cite{Israel:1967za}. 
We also find another useful equation 
\begin{align}
\label{}
D^2 G_2 -\frac 1V D^i VD_i G_2=0 \,, \qquad 
G_2(V, \psi) \equiv V^2 -(\psi-\psi_1)^2 \,,
\end{align}
where $\psi_1 $ is an arbitrary constant. If one chooses $\psi_1=(1+\psi_0^2)/(2\psi_0)$, 
the values of $G_2$ at infinity and at horizon are equal. By the maximum/minimum principle,  
$G_2(V, \psi)$ is constant,  yielding (\ref{RN_psiV}).

Since the functional dependence of $\psi $ on $V $ has been specified, 
the field equations are now simplified to 
\begin{align}
\label{}
D^2 V=\frac{q^2V}{\beta(V)^2}(DV)^2 \,, 
\end{align}
and 
\begin{align}
\label{}
{}^{(3)}R_{ij}=\frac 1VD_iD_jV+\frac{q^2}{\beta(V)^2}[(DV)^2 g_{ij}-2D_iVD_jV] \,.   
\end{align}

As in the vacuum case, let us define a three dimensional symmetric tensor
\begin{align}
\label{}
H_{ij} \equiv D_iD_jV+\frac{V}{1-V^2}\frac{(\beta(V)+1)(4\beta(V)-1)}{\beta(V)^2}D_iVD_jV
-\frac{V(1+\beta(V))}{(1-V^2)\beta(V)} (DV)^2 g_{ij}\,, 
\end{align}
and a vector field
\begin{align}
\label{}
H_i \equiv \frac{D_i \rho }{\rho }-\frac{V(\beta(V)+1)(3\beta(V)-1)}{(1-V^2)\beta(V)^2}D_iV\,. 
\end{align}
These tensors  satisfy the relations identical to the vacuum case
\begin{align}
\label{evHij}
H_{ij}D^j V=- \rho^{-2} H_i \,, \qquad H^i{}_i=0\,.
\end{align}
When $Q=0$, these tensors reduce to (\ref{vac_Hij}) and (\ref{vacHi}). 
In terms of local quantities of $\ma S_V$, we have 
\begin{align}
\label{}
H_{ij}=\rho^{-1}\sigma_{ij} -2 \rho^{-2}n_{(i}\ma D_{j)} \rho +\frac{1}{2\rho}
\left[k -\frac{2V(1+\beta(V))}{\rho \beta(V)(1-V^2)}\right](h_{ij}-2n_in_j)\,. 
\end{align}
Hence, $H_{ij}=0$ is a condition for the spherical symmetry. This will be also apparent by 
looking at the Cotton tensor $C_{ijk}=2D_{[i}[{}^{(3)}R_{j]k}-({}^{(3)}R/4)g_{j]k}]$, which is 
given in terms of $H_{ij}$ as 
\begin{align}
\label{}
C_{ijk}=\frac 2{V^2}\left(1-\frac{q^2V^2}{\beta(V)^2}\right)
(2H_{k[i}D_{j]}V-\rho^{-2}g_{k[i}H_{j]}) \,. 
\end{align}
It then suffices to show $H_{ij}=0$ for the uniqueness of the Reissner-Nordstr\"om solution (\ref{RN}).

We wish to find the current conservation equation of the separable type (\ref{currentcons}). 
Using 
\begin{align}
\label{}
D^2 \rho=-\rho^3|D_iD_jV|^2+\frac 3{\rho}(D \rho)^2+D^i\rho D_i V \left(\frac 1V
+\frac{2q^2V}{\beta(V)^2}\right)+\frac{2q^4V^2}{\rho \beta(V)^4} \,,   
\end{align}
the divergence of $J^i$ culminates in a separable form (\ref{ansatz}) with respect to $\rho $ and $V$, if we choose
\begin{align}
\label{}
g_1(\rho) =-c \rho^{-(c+1)} \,, \qquad g_2 (\rho )= \rho^{-c} \,, 
\end{align}
where $c $ is a constant. With this choice, we are left with two linear first-order equations
which are easily solved to give 
\begin{align}
\label{}
f_1(V)=&\frac{1}{V(1-V^2)^{2c-1} } \frac{[1+\beta(V)]^{2c-2}}{\beta(V)^{c}}\bigl[a(1+\beta(V))+2b (1-V^2)\bigr]\,, \\ 
f_2(V)=&\frac{[1+\beta(V)]^{2c-1}}{(1-V^2)^{2c} \beta(V)^{2+c}}\Bigl[
-a \beta(V)(1+\beta(V))+c(3\beta(V)-1)\{a(1+\beta(V))+2b(1-V^2)\}
\Bigr]\,,  
\end{align}
where $a$ and $b$ are integration constants. We thus conclude that the following divergence equation holds 
\begin{align}
\label{RN_divJ}
D_i J^i =\frac{cf_1(V)}{2\rho^c}\left[
\left|2\rho^2 D_{[i}VH_{j]k}-H_{[i}g_{j]k}\right|^2 
+\left(2c-1 \right)|H_i|^2 
\right]\,. 
\end{align}
This is of the same form as in the vacuum case (\ref{vac_divid}). 
The right hand side of this equation becomes nonnegative, provided
\begin{align}
\label{EMineqcab}
c\ge \frac 12 \,, \qquad a \ge 0 \,, \qquad a+\frac{2}{1+\sqrt{1-q^2}} b\ge 0 \,. 
\end{align}
The last two conditions ensure $f_1(V)\ge 0$. 

By making use of Stokes' theorem and inserting the boundary conditions 
(\ref{EV_BCinf}), (\ref{EV_BCH}) with (\ref{EV_trGausseq}), one gets 
\begin{align}
\label{Uni_RN_Fint}
%0\le  &-4 \pi a M^{1-c} 
%+2 c \pi \chi (1-q^2)^{-c/2} (1+\sqrt{1-q^2})^{2(c-1)} (a-b+a\sqrt{1-q^2}) \rho_0^{1-c} 
%\notag \\
%&+A_H(1-q^2)^{-c/2-1}(1+\sqrt{1-q^2})^{2c}
%[a\sqrt{1-q^2}+c\{b-a(1+\sqrt{1-q^2})\}]\rho_0^{-(1+c)}\,.
0\le  & a \left[-4 \pi  M^{1-c} +A_H(1-q^2)^{-(c+1)/2}(1+\sqrt{1-q^2})^{2c}\rho_0^{-(1+c)} \right] \notag \\
&+\left[ a (1+ \sqrt{1-q^2})+2 b \right] c (1-q^2)^{-c/2} (1+\sqrt{1-q^2})^{2(c-1)}\rho_0^{1-c}  \notag \\
& \hspace{2cm}
\times \left[ 2\pi \chi - A_H(1-q^2)^{-1}(1+\sqrt{1-q^2})^{2}\rho_0^{-2} \right]\,.
\end{align}
Here, $\chi$ is the Euler characteristic (\ref{Euler}) of the horizon cross-section. 
The  necessary and sufficient condition for which this inequality holds for any values of $a$, $b$ and $c$ satisfying (\ref{EMineqcab}) is
\begin{align}
\label{}
4 \pi \left( \frac{M(1+\sqrt{1-q^2})^2}{\rho_0 \sqrt{1-q^2}}  \right)^{1-c} \le A_H \left( \frac{(1+\sqrt{1-q^2})}{\rho_0\sqrt{1-q^2}}\right)^2  \le  2 \pi \chi \,.
\end{align}

The case for $c=1$ immediately gives $\chi \ge 2$, and hence we set $\chi=2$ in the hereafter. 
Then this inequality for $c=1$ is consistent only if the equality holds, that is
\begin{align}
\label{Uni_RN_ineq1}
A_H = \frac{4\pi \rho_0^2(1-q^2)}{(1+\sqrt{1-q^2})^2} \, .
\end{align}
On account of $H_{ij}=0$, the spacetime is spherically symmetric and therefore 
the Reissner-Nordstr\"om solution (\ref{RN}) is singled out. 

Note that we did use Smarr's formula (\ref{Smarr:EV}) to derive (\ref{RN_psiV}), 
in contrast to the vacuum case. Nonetheless, we have nowhere used the property of the symmetric coset space, in contrast to the argument in \cite{Simon1984}. 
In this sense, the present proof seems to have a potentially broader applicability.

\section{Einstein-Maxwell-dilaton theory}
\label{sec:EMD}

We next consider the four-dimensional Einstein-Maxwell-dilaton gravity described by Lagrangian
\begin{align}
\label{EMD_action}
L= R-2g^{\mu\nu}\partial_\mu \phi \partial_\nu \phi -e^{-2\alpha \phi}F_{\mu\nu}F^{\mu\nu} \,, 
\end{align}
where $\alpha(>0)$ is a coupling constant. 
Let us focus on the static spacetimes, where the metric (\ref{metric}), 
the Maxwell field and the dilaton field are all invariant under the orbit of static Killing field. 
Here we assume that the Maxwell field is electric 
$F=-\D \psi \we \D t$, where $\psi$ is the $t$-independent electrostatic potential. 
Note that the absence of the magnetic potential is a genuine restriction, 
since the equations of motion arising from the action (\ref{EMD_action}) do not admit a duality rotation. 
Under these conditions, Einstein's equations are simplified to 
\begin{align}
\label{}
D^2 V&=\frac 1V e^{-2\alpha \phi}(D \psi)^2 \,, \\
{}^{(3)}R_{ij}&=\frac 1VD_i D_j V-\frac 2{V^2}e^{-2\alpha \phi} 
\left(D_i \psi D _j\psi -\frac 12 g_{ij}(D \psi)^2\right)
+2D_i\phi D_j \phi \,. 
\end{align}
The scalar curvature of the three-dimensional space reads
${}^{(3)}R=2V^{-2}e^{-2\alpha\phi}(D \psi)^2+2(D \phi)^2$. 
The Maxwell and dilaton equations of motion are given by 
\begin{align}
\label{EMD_Maxwell}
D_i(e^{-2\alpha \phi}V^{-1}D^i \psi)&=0 \,, \\
D_i(VD^i\phi)-\frac{\alpha}{V}e^{-2\alpha \phi} (D \psi)^2 &=0 \,. 
\label{EMD_dilaton}
\end{align}
The finiteness of curvature invariants 
at the horizon requires
\begin{align}
\label{}
k_{ij}|_{V=0}=0 \,, \qquad \rho|_{V=0}=\rho_0 \,, \qquad 
\psi|_{V=0}= \psi_0\,, \qquad  \textrm{$\phi$, $\partial_V\phi$, $\ma D_i\phi$ are finite}\,, 
\end{align}
where $\rho_0$ ($0<\rho_0<\infty$) and $\psi_0$ are constants, representing the inverse of the surface gravity 
and the electrostatic potential at the horizon. Let us emphasize that the value of the dilaton field 
is not necessarily constant over the horizon, i.e., $\ma D_i\phi|_{V=0}\ne 0$ in general.  

The boundary conditions at infinity  are 
\begin{align}
\label{}
g_{ij}\sim \left(1+\frac{2M}{r}\right)\delta_{ij} \,, \qquad 
V\sim 1-\frac{M}{r} \,, \qquad \psi\sim \frac{Q}{r}\,, \qquad 
%\phi \sim -\frac{\alpha Q^2}{(M+\sqrt{M^2-Q^2(1-\alpha^2)})r}\,. 
\phi \sim \frac{\phi_\infty}{r} \,.
\end{align}
Here $M$ is the ADM mass, $Q$ is the electric charge and $\phi_\infty$ is a constant.  
From the positive mass theorem~\cite{Gibbons:1993xt,Nozawa:2010rf}, the mass 
is bounded from below by the charge as
\begin{align}
\label{BPSdilaton}
M\ge \frac{Q}{\sqrt{1+\alpha^2}} \,. 
\end{align}
From now on, we assume that this inequality is strictly satisfied.

A static and spherically symmetric black-hole solution to this theory was discovered by Gibbons and Maeda \cite{Gibbons:1987ps}, whose metric boils down to 
\begin{align}
\label{GMsol}
\D s^2=- \left(1-\frac{r_+}{r}\right)\left(1-\frac{r_-}{r}\right)^{\frac{1-\alpha^2}{1+\alpha^2}}\D t^2+
\left(1-\frac{r_+}{r}\right)^{-1}\left(1-\frac{r_-}{r}\right)^{-\frac{1-\alpha^2}{1+\alpha^2}}\D r^2+r^2\left(1-\frac{r_-}{r}\right)^{\frac{2\alpha^2}{1+\alpha^2}}  \D \Omega_2^2 \,, 
\end{align}
with
\begin{align}
\label{}
\phi=\frac{\alpha }{1+\alpha^2}\log \left(1-\frac{r_-}{r}\right)\,, \qquad 
\psi=  \frac{\sqrt{r_+r_-}}{\sqrt{1+\alpha^2 }r }  \,. 
\end{align}
Here $r_+(>r_->0)$ is a locus of the event horizon and $r_-$ corresponds to the curvature singularity.
These two parameters are related to the ADM mass $M$ and the electric charge $Q$ by 
\begin{align}
\label{rpm}
 r_\pm =\frac{1+\alpha^2 }{1\pm \alpha^2 } 
\left(M\pm \sqrt{M^2-(1-\alpha^2)Q^2 }\right)\,. 
\end{align}
The uniqueness property of this solution was first addressed in \cite{MuA:dilaton} 
for a coupling constant $\alpha=1$,  by finding a suitable Witten spinor on a spatial timeslice. 
Subsequently, \cite{Mars:2001pz} extended the proof to encompass the case of arbitrary $\alpha$ using the conformal positive mass theorem. These techniques were also worked out in higher dimensions \cite{Gibbons:2002av,Gibbons:2002ju}. In the context of divergence equations, 
the most difficult issue is how to assess the value of the dilaton field at the event horizon. 
In this section, we overcome this obstacle by finding yet another divergence equation. 
We develop a new proof based only upon divergence equations.

One can rewrite some of the equations of motion into the divergence type 
\begin{align}
\label{EMD_uni_divid1}
D_i \left(D^iV-\frac 1V e^{-2\alpha\phi}\psi D^i\psi\right)&=0 \,, \\
\label{EMD_uni_divid2}
D_i \left[\frac 1V \{(1+\alpha^2)e^{-2\alpha \phi}\psi ^2 +V^2 \}D^i \psi -2\psi D^i V 
-2 \alpha  V\psi D^i \phi \right]&=0 \,, \\
\label{EMD_uni_divid3}
D_i \left(VD^i\phi-\alpha D^i V\right)&=0 \,.
\end{align}
Integration of (\ref{EMD_Maxwell}), (\ref{EMD_uni_divid1}), (\ref{EMD_uni_divid2}) and (\ref{EMD_uni_divid3}) gives 
\begin{align}
\label{}
4\pi M-A_H \rho_0^{-1}+\frac{\psi_0}{\rho_0} \int _B e^{-2\alpha \phi }\frac{\partial_V \psi}{V} \D S&=0 \,, \\
4\pi Q+\frac{1}{\rho_0} \int _B e^{-2\alpha \phi }\frac{\partial_V \psi}{V} \D S&=0\,, \\
-4\pi Q+2 \psi_0 \rho_0^{-1} A_H-\frac{1+\alpha^2}{\rho_0} \psi_0^2 \int _B e^{-2\alpha \phi }\frac{\partial_V \psi}{V} \D S&=0\,,\\
-4 \pi \phi_\infty -4\pi \alpha M +\alpha A_H \rho_0^{-1} &=0 \,.
\end{align}
These equations are combined to give 
\begin{align}
\label{}
A_H=4\pi \rho_0(M-Q \psi_0)\,, \qquad \psi_0=\frac{Q}{M+\sqrt{M^2-Q^2(1-\alpha^2)}} \,, \qquad 
\phi_\infty = -\alpha Q \psi_0 \,.
\end{align}
It follows that $\phi_\infty$ is a secondary charge, 
since it is specified by $M$ and $Q$. 

To proceed, let us define~\cite{Mars:2001pz}
\begin{align}
\Phi_{\pm 1}= &\frac 12 \left[
e^{\alpha \phi}V\pm \frac{e^{-\alpha \phi}}{V}-(1+\alpha^2)\frac{e^{-\alpha \phi}\psi^2}{V}
\right]  \,, \\
\Phi_0=& \sqrt{1+\alpha^2} \frac{e^{-\alpha \phi}\psi }{V} \,, \\
\Psi_{\pm 1}=& \frac 12 \left(e^{-\phi /\alpha }V\pm e^{\phi /\alpha }V^{-1} \right)\,, 
\end{align}
satisfying the constraints
\begin{align}
\label{EMD_PhiPsiconst}
\Phi_{+1}^2-\Phi_{-1}^2+\Phi_{0}^2=1\,, \qquad \Psi_{+1}^2- \Psi_{-1}^2 = 1 \,. 
\end{align}
These scalars span the homogeneous coordinates of 
${\rm SO}(1,1) \times {\rm SL}(2,\mathbb R)/{\rm SO}(1,1)$, which 
corresponds to the nonlinear sigma model of static Einstein-Maxwell-dilaton theory. 
The equations of motions for these scalars read
\begin{align}
\label{}
D_i(VD^i \Phi_{-1/0/+1})= V G^{(\Phi)} \Phi_{-1/0/+1} \,, \qquad 
D_i(VD^i \Psi_{\pm 1})= V G^{(\Psi)}  \Psi_{\pm 1}\,, 
\end{align}
where 
\begin{align}
\label{}
G^{(\Phi)}=|D_i \Phi_{-1}|^2  -|D_i \Phi_{0}|^2 -|D_i \Phi_{+1}|^2 \,, \qquad 
G^{(\Psi)}= |D_i\Psi_{-1}|^2-|D_i\Psi_{+1}|^2\,. 
\end{align}

From the field equations, 
the scalars $G_\pm =1-Ve^{\alpha \phi}\pm \sqrt{1+\alpha^2}\psi $ obey 
\begin{subequations}
\begin{align}
\label{}
%D^2 F_\pm + \left(
%\mp \sqrt{1+\alpha^2}\frac{e^{-\alpha\phi}}{V}D_i \psi -\alpha D_i \phi
%\right) D^i F_\pm &=0\,, \\
D^2G_\pm+ \left(
\pm \sqrt{1+\alpha^2}\frac{e^{-\alpha\phi}}{V}D_i \psi -\alpha D_i \phi
\right) D^i G_\pm &=0\,. 
\end{align}
\end{subequations}
These are elliptic differential equations for which the maximum/minimum principle can be applied~\cite{elliptic}. 
Since $G_\pm \to 0$ at infinity and $G_\pm \to (1\pm \sqrt{1+\alpha^2}\psi_0)>0$ at the horizon, 
we see that $G_\pm$ never attains zero inside $\Sigma$. 
We therefore have 
$\Phi_{+1}-1=\frac 1{2}V^{-1}e^{-\alpha \phi}G_+G_- >0$ inside $\Sigma$. 

Similarly, 
we obtain
\begin{align}
\label{}
D^2 \left(\frac{e^{\phi/\alpha}}{V}-1\right)+D^i \left(2\log V-\frac{\phi}{\alpha}\right)
D_i \left(\frac{e^{\phi/\alpha}}{V}-1\right)=0\,. 
\end{align}
By virtue of $e^{\phi/\alpha}-V\to 0$ at infinity and $e^{\phi/\alpha}-V\to e^{\phi/\alpha}$ at the event horizon, 
the minimum of $V^{-1}e^{\phi/\alpha}-1$ must be attained at infinity, and then  in the interior of  $\Sigma$ we have 
$e^{\phi/\alpha}-V> 0$. 
This means that 
$\Psi_{+1}-1=\frac 12V^{-1}e^{-\phi/\alpha}(e^{\phi/\alpha}-V)^2$ is strictly positive  in the interior of  $\Sigma$.  

Since the conditions $\Phi_{+1},\Psi_{+1}>1$ have been demonstrated, 
we now move on to show that two of the inhomogeneous coordinates ($V, \phi, \psi$) of 
${\rm SO}(1,1)\times {\rm SL}(2,\mathbb R)/{\rm SO}(1,1)$ are linearly dependent. 
In particular, our current aim is to demonstrate that 
$\Phi_{+1}/\Psi_{+1}$ and $\Phi_0^2/(\Phi_{+1}^2-1)$ are constants. 
Guided by the argument to derive (\ref{vac_divid}), 
we consider the divergence equation with the following current 
\begin{align}
\label{}
J^i= V[f_1(\Phi_{+1})g_1(\Psi_{+1}) D^i \Phi_{+1}+f_2(\Phi_{+1})g_2(\Psi_{+1}) D^i \Psi_{+1} ]\,.  
\end{align}
By choosing 
\begin{align}
\label{}
f_1(\Phi_{+1})&=\frac{(\Phi_{+1}+\sqrt{\Phi_{+1}^2-1})^{-c}}{\sqrt{\Phi_{+1}^2-1}}\,, \qquad 
f_2(\Phi_{+1})=-(\Phi_{+1}+\sqrt{\Phi_{+1}^2-1})^{-c} \,, \notag \\
g_1(\Psi_{+1})&=(\Psi_{+1}+\sqrt{\Psi_{+1}^2-1})^c\,, \qquad 
g_2(\Psi_{+1})=\frac{(\Psi_{+1}+\sqrt{\Psi_{+1}^2-1})^c}{\sqrt{\Psi_{+1}^2-1}} \,, 
\end{align}
where $c$ is a constant, 
some amount of algebra shows that the following equation holds
\begin{align}
D_i J^i=& -V\left(\frac{\Psi_{+1}+\sqrt{\Psi_{+1}^2-1}}{\Phi_{+1}+\sqrt{\Phi_{+1}^2-1}}\right)^c
\Biggl[\frac{\Phi_{+1}\sqrt{\Phi_{+1}^2-1}}
{\Phi_{+1}^2+\Phi_{0}^2-1}
\left|D_i \Phi_{0}-\frac{\Phi_{0}\Phi_{+1}}{\Phi_{+1}^2-1}D_i \Phi_{+1}\right|^2\notag \\
& +\frac{c}{\Psi_{+1}^2-1}\left|D_i \Psi_{+1}-\sqrt{\frac{\Psi_{+1}^2-1}{\Phi_{+1}^2-1}}D_i \Phi_{+1}\right|^2 \Biggr] \,.
\label{EMD_dividPhiPsi}
\end{align}
For $c>0$, the right-hand side of this equation becomes negative semi-definite. 
We can thus derive the inequality
\begin{align}
\label{}
\int_\Sigma D_i J^i \D \Sigma =\int _{S^\infty} J^i n_i \D S-\int _{B} J^i n_i \D S \le 0 \,. 
\end{align}
Our present boundary conditions tell us that both of the surface integrals at infinity and at horizon vanish, 
regardless of the value of $c(>0)$. 
This means that the equality is saturated and the each piece on the right-hand side of 
(\ref{EMD_dividPhiPsi}) vanishes independently, leading to 
\begin{align}
\label{}
\frac{\Phi_{+1}}{\Psi_{+1}}=1 \,, \qquad 
\frac{\Phi_0^2}{\Phi_{+1}^2-1}=\frac{4(1+\alpha^2)\psi_0^2}{[(1+\alpha^2)\psi_0^2-1]^2} \,,
\end{align}
where the integration constants have been fixed by asymptotic value. 
These equations give 
\begin{align}
\label{EMD_Vpsi}
V^2=\frac{e^{2\phi/\alpha}-[1-(1+\alpha^2)\psi_0^2]e^{\frac{1-\alpha^2}{\alpha}\phi}}{(1+\alpha^2)\psi_0^2}\,, \qquad 
\psi=\frac{1-e^{\frac{1+\alpha^2}{\alpha}\phi}}{(1+\alpha^2)\psi_0} \,. 
\end{align}
It is worth commenting that the value of the dilaton field at the horizon is constant 
\begin{align}
\label{}
\phi|_{V=0}= \frac{\alpha}{1+\alpha^2}\log[1-(1+\alpha^2)\psi_0^2] \,. 
\end{align}

Since the functional dependence of $(V, \phi, \psi)$ has been established, 
our remaining task is to show the spherical symmetry following the prescription given in the foregoing sections. 
At this final step of the proof, 
it is of advantage to consider the conformal transformation 
\begin{align}
\label{}
g_{ij}=e^{2\alpha \phi} \hat  g_{ij} \,, \qquad 
\hat V=V e^{\alpha \phi} \,, \qquad \hat \psi=\sqrt{1+\alpha^2} \psi \,. 
\end{align}
With the relation (\ref{EMD_Vpsi}), one can check that the following relations are satisfied
\begin{align}
\label{}
&\hat R_{ij}= \frac{1}{\hat V}\hat D_i\hat D_j \hat V-\frac 2{\hat V^2}
\left(\hat D_i \hat\psi \hat D_j\hat\psi -\frac 12 \hat g_{ij}(\hat D\hat\psi)^2 \right) \,, \nonumber \\
& \hat D_i (\hat V^{-1} \hat D^i \hat \psi ) =0 \,, \qquad 
\frac{1-\hat V^2}{\hat \psi}+\hat \psi =\hat \psi_0^{-1}+\hat \psi_0 \,. 
\end{align}
In terms of these new variables, 
the boundary conditions at infinity reduce to 
\begin{align}
\label{}
\hat V \sim 1-\frac{r_++r_-}{2r} \,, \qquad \hat g_{ij} \sim \left(1+\frac{r_++r_-}{r}\right)\delta_{ij}\,,\qquad 
\hat \psi \sim \frac{\sqrt{r_+r_-}}r \,, 
\end{align}
where $r_\pm$ is given by (\ref{rpm}). The event horizon is located at $\hat V=0$ where $\hat \psi$ 
and $(\hat D\hat V)^2$ are constants. 
Since these conditions are exactly the same as in the electrovacuum case in section \ref{sec:electrovac},  
the solution must be spherical, i.e.,  
\begin{align}
\label{}
\hat V^2=\left(1-\frac{r_+}{r}\right)\left(1-\frac{r_-}{r}\right)\,, \qquad 
\hat g_{ij}\D x^i\D x^j = \frac{\D r^2}{\hat V^2} +r^2 \D \Omega_2^2 \,, \qquad 
\hat \psi=\frac{\sqrt{r_+r_-}}r\,, 
\end{align}
and hence $e^{(1+\alpha^2)\phi/\alpha}=1-\hat \psi\hat \psi_0=1-r_-/r$. 
Going back to the original frame by $g_{ij}=e^{2\alpha \phi} \hat  g_{ij}$, 
we recover the Gibbons-Maeda solution (\ref{GMsol}).

%Furthermore, we have the following equation
%\begin{align}
%\label{}
%D^2 \left[
%\frac{1-e^{2\alpha \phi}V^2}{\psi}+(1+\alpha^2)\psi
%\right]-D^i \left[\log \left(\frac{V e^{2\alpha\phi}}{\psi^2}\right)\right]
%D_i \left[
%\frac{1-e^{2\alpha \phi}V^2}{\psi}+(1+\alpha^2)\psi
%\right]=0 \,. 
%\end{align}
%Inserting the present boundary conditions  
%one sees that the value of $\psi^{-1}(1-e^{2\alpha \phi}V^2)+(1+\alpha^2)\psi$
%at the horizon and at infinity coincides. 
%Applying the maximum principle, $\psi^{-1}(1-e^{2\alpha \phi}V^2)+(1+\alpha^2)\psi$ is constant, i.e.,   
%\begin{align}
%\label{}
%\psi =\frac{1+(1+\alpha^2)\psi_0^2 - \sqrt{[(1+\alpha^2)\psi_0^2-1]^2+4(1+\alpha^2)\psi_0^2 e^{2\alpha \phi}V^2 }}
%{2(1+\alpha^2)\psi_0} \,. 
%\end{align}
%The minus sign in front of the square root has been chosen due to the relation 
%$(1+\alpha^2)\psi_0^2-1<0$, which follows from the Bogomol'ny bound (\ref{BPSdilaton}). 

\section{Higher dimensions}
\label{sec:higherD}

It is natural to inquire whether our algorithm is applicable in higher dimensions. 
For simplicity, 
we focus on solutions to the $n$-dimensional vacuum Einstein equations 
$R_{\mu\nu}=0$, which reduce in the static spacetime 
$\D s^2=-V^2(x)\D t^2 +g_{ij}(x) \D x^i \D x^j $ to
\begin{align}
\label{}
D^2 V=0 \,, \qquad {}^{(n-1)} R_{ij}=\frac 1VD_iD_j V\,,
\end{align}
where $ {}^{(n-1)} R_{ij}$ is the Ricci tensor for the ($n-1$)-dimensional 
spatial metric $g_{ij}$. 
Assuming the local foliation $\ma S_V=\{V={\rm constant}\}$ of constant timeslice $\Sigma$, the following relations are
satisfied
\begin{align}
\label{HigherEineq_geometric}
n^i D_i \rho =\rho k \,, \qquad {}^{(n-2)}R =\frac{2k}{V\rho}+k^2-k_{ij}k^{ij} \,. 
\end{align}
Here $n_i =\rho D_i V$ is the outward pointing unit normal to $\ma S_V$ in $\Sigma$ with 
$\rho $ being the lapse function and we shall denote the induced metric of $\ma S_V$
as $h_{ij}=g_{ij}-n_in_j$ as before. 
$k_{ij}$ is the extrinsic curvature of $\ma S_V$ in $\Sigma$ and it splits up into the trace-free part and the trace part
\begin{align}
\label{}
k_{ij} =\sigma_{ij} + \frac {k}{n-2} h_{ij}  \,\,, \qquad k= h^{ij} k_{ij}.  
\end{align}

The boundary conditions at the horizon are
\begin{align}
\label{}
\left.k_{ij}\right|_{V=0}=0\,, \qquad 
\rho|_{V=0}=\rho_0\,,  
\end{align}
where $\rho_0$ is a positive constant. 
The asymptotic flatness is
\begin{align}
\label{higherd_asyBC}
V\sim 1-\frac{m}{r^{n-3}} \,, \qquad 
g_{ij}\sim \left(1+\frac{2m}{(n-3)r^{n-3}}\right) \delta_{ij}\,, 
\end{align}
where $m$ corresponds to the ADM mass up to a constant. 
In terms of lapse and mean curvature, (\ref{higherd_asyBC}) is translated into
\begin{align}
\label{}
\rho \sim\frac{r^{n-2}}{(n-3)m}\,, \qquad k\sim \frac{n-2}{r} \,. 
\end{align}

The $(n-1)$-dimensional tensor quantities which we wish to show them to vanish 
are 
\begin{align}
\label{Hij_higher}
H_{ij}= D_i D_j V-\frac 2{n-3}\frac{V(DV)^2}{1-V^2}g_{ij}+\frac{2(n-1)}{n-3}\frac{V}{1-V^2}D_i V D_j V  \,,
\end{align}
and 
\begin{align}
\label{}
H_i =\frac{D_i\rho}{\rho }-\frac{2(n-2)}{n-3}\frac{V}{1-V^2}D_i V\,. 
\end{align}
These quantities satisfy
\begin{align}
\label{}
H_{ij}D^j V=-\rho^{-2}H_i \,, \qquad H^i{}_i=0 \,. 
\end{align}
In the geometric notation using the data on $\mathcal{S}_V$, we have
\begin{align}
\label{Sch_Uni_Hmndechigh}
 H_{ij}=&\rho^{-1}\sigma_{ij}-\frac 2{\rho^2}n_{(i}\ma D_{j)}\rho +\frac 1{(n-2)\rho}[h_{ij}-(n-2)n_in_j]
\left(k-\frac{2(n-2)}{n-3}\frac{V}{\rho(1-V^2)}\right) \,. 
\end{align}
Let us now demonstrate that $H_{ij}=0$ indeed implies that the space is spherically symmetric, 
i.e., it admits an isometry of ${\rm SO}(n-1)$. 
Suppose that $H_{ij}=0$ holds. Then $H_i=0$ is readily solved to give 
\begin{align}
\label{rhoVhigher}
[(n-3)\rho]^{n-3} \left(\frac{1-V^2}{2}\right)^{n-2}=m \,, 
\end{align}
where the integration constant has been determined by the asymptotic condition. 
Next, $\sigma_{ij}=0$ implies that $k_{ij}=\frac 1{2\rho}\partial_V h_{ij}=\frac{1}{n-2}k h_{ij}$. 
Integrating this equation by use of (\ref{rhoVhigher}), we are led to
$h_{ij}=[(n-3)m \rho]^{2/(n-2)} \hat h_{ij}$, where $\hat h_{ij}$ is a metric which is independent of $V$. 
From the vacuum Einstein equations, $\hat h_{ij}$ is the Einstein metric of positive curvature. 
The asymptotic flatness requires that this must be a standard metric of a unit sphere. 
It follows that $H_{ij}$ represents a deviation from the spherical symmetry also in higher dimensions.

We shall not attempt to derive in detail the divergence equation, but 
only show the final results, since the procedure is completely parallel with 
the four dimensional case. 
Starting with the separable ansatz (\ref{ansatz}), we have a higher dimensional version of the 
divergence equation 
\begin{align}
\label{higherD_divJ}
D_i J^i= \frac{(n-3)c}{2\rho^c}f_1(V) \left[
\left|2\rho^2D_{[i}VH_{j]k}-\frac{2}{n-2}H_{[i}g_{j]k} \right|^2
+2\left(c-\frac{n-3}{n-2}\right) |H_i|^2 
\right]\,, 
\end{align}
where $g_1(\rho)=-(n-3)c\rho^{-(1+c)}$, $g_2(\rho)=\rho^{-c}$ and 
\begin{align}
\label{}
f_1(V)=&\frac{1}{V(1-V^2)^{\frac{n-2}{n-3}c-1}} [a+b(1-V^2)] \,, \notag \\
f_2(V)=&\frac{2}{(1-V^2)^{\frac{n-2}{n-3}c}} \left[
a\{c(n-2)-(n-3)\}+bc(n-2)(1-V^2) 
\right] \,.
\end{align}
Alternatively, the current $J^i$  can be again put into a more useful form 
\begin{align}
\label{}
J^i=-(n-3) [(1-V^2)^{\frac{n-2}{n-3}}\rho ]^{-c} \left[\frac{c}{V} (1-V^2)[a+b(1-V^2)]H^i +2aD^i V\right] \,. 
\end{align}
The right-hand side of (\ref{higherD_divJ}) becomes positive semi-definite if 
\begin{align}
\label{higherD_para}
a\ge 0 \,, \qquad 
a+b\ge 0 \,, \qquad 
c\ge \frac{n-3}{n-2} \,. 
\end{align}

It is worth commenting that the right-hand side of (\ref{higherD_divJ}) is expressed 
by means of the tensor $H_{ij}$ only. This term is not expressible 
by the higher dimensional Cotton tensor 
$C_{ijk}=2 D_{[i}({}^{(n-1)}R_{j]k}-\frac 1{2(n-2)}{}^{(n-1)}Rg_{j]k})$
 (note that this is not conformally invariant unless $n=4$),
since the Weyl tensor $ {}^{(n-1)}\!C_{ijkl}$ of $(\Sigma, g_{ij})$  becomes relevant as 
\begin{align}
\label{}
C_{ijk}=
 {}^{(n-1)}\!C_{ijkl}\frac{D^lV}V
-\frac{2}{(n-3)V^2}[(n-2)D_{[i}{V}H_{j]k}-\rho^{-2}H_{[i}g_{j]k} ]\,.
\label{higherD_Cotton}
\end{align}

\subsection{Surface integral}

As discussed above, the uniqueness of the Schwarzschild black hole follows, provided 
that one can show $H_{ij}=0$ under our boundary conditions. 
The integration of (\ref{higherD_divJ}) over the spatial slice $\Sigma$ 
boils down to 
\begin{align}
\label{higher_surfaceint}
%0\le &-2^{-c}a (n-3)^{c+2}(2m)^{1-\frac{c}{n-3}} \Omega_{n-2} 
%+2A_H \rho_0^{-(1+c)}\left[
%(n-3)a -c(n-2)(a+b)
%\right]\notag \\
%&+\frac 12 (a+b)c (n-3) \rho_0^{1-c} \int _B {}^{(n-2)}\!R \D S \,. 
0\le &2a(n-3)\left[-\left(\frac {(n-3)} 2\right)^{c+1}(2m)^{1-\frac{c}{n-3}} \Omega_{n-2} 
+A_H \rho_0^{-(1+c)}\right] \notag \\
& +c(a+b)\rho_0^{1-c} \left[
\frac {(n-3)}2 \int _B {}^{(n-2)}\!R \D S- 2 (n-2) A_H \rho_0^{-2}
\right] \,. 
\end{align}
This inequality holds for any values of $a$, $b$ and $c$ satisfying (\ref{higherD_para}), if and only if the pair of inequalities
\begin{align}
\label{higherDineq}
\left(\frac  {(n-3) } 2 \rho_0 \right)^{c+1}(2m)^{1-\frac{c}{n-3}} \Omega_{n-2} \le 
A_H  \le \frac{(n-3)\rho_0^2}{4(n-2)}\int_B {}^{(n-2)}R \D S \,
\end{align}
is satisfied.

%\begin{align}
%\label{AHlower_higher}
%\Omega_{n-2} \left(\frac{n-3}{2}\rho_0\right)^{n-2}\le A_H \,. 
%\end{align}
Combining the former inequality with the Smarr relation
\begin{align}
\label{Smarr_higher}
(n-3)m \Omega_{n-2}\rho_0=A_H \,, 
\end{align}
for any $c$, one obtains the Penrose-type inequality
\begin{align}
\label{Penroseineq_higher}
A_H \le \Omega_{n-2}(2m)^{\frac{n-2}{n-3}} \,. 
\end{align}
We wish to show that, in (\ref{higherDineq}), the at-most-right-hand side coincides with 
the at-most-left-hand side, which then results in the equalities.
Unfortunately, the value of $\int_B {}^{(n-2)}R \D S$ cannot be evaluated in higher dimensions in general, 
since it is not a topological invariant. 
Only the lower bound of $\int_B {}^{(n-2)}R \D S$  is obtainable.  
For instance, the case of $c=n-3$ gives
rise to a lower bound 
\begin{align}
\label{}
(n-2)(n-3)^{n-3}\Omega_{n-2} \le \left(\frac{2}{\rho_0}\right)^{n-4} \int_B {}^{(n-2)}R \D S\,. 
\end{align}
%The inequality (\ref{higherDineq}) only gives the lower bound. 

To obtain further insight, let us rewrite the second inequality in (\ref{higherDineq}) into a more 
recognizable form. 
For this purpose, let us define the analogue of the Yamabe constant by
\begin{align}
\label{yamabe}
y_H\equiv \frac{Y_H}{Y_H^0}\,, \qquad 
Y_H \equiv \frac{\int _B {}^{(n-2)}R\D S }{A_H^{\frac{n-4}{n-2}}} \,, \qquad 
Y_H^0\equiv (n-2)(n-3)\Omega_{n-2}^{2/(n-2)} \,. 
\end{align}
In terms of $y_H$, 
the Smarr relation (\ref{Smarr_higher}) recasts the latter inequality of (\ref{higherDineq}) into (note that the exponent of $y_H$ is different from
\cite{Mizuno:2009fj})
\begin{align}
\label{revPenroseineq_higher}
\Omega_{n-2}(2m)^{\frac{n-2}{n-3}}\le A_H y_H^{\frac{n-2}{2(n-3)}}\,. 
\end{align}
If one can show
\begin{align}
\label{yHbound}
y_H \le 1\,,
\end{align}
inequalities (\ref{Penroseineq_higher}) and (\ref{revPenroseineq_higher}) are consistent only if $y_H=1$, 
yielding spherical symmetry. 
Recall that the currently only proof of the uniqueness of the higher dimensional Schwarzschild solution  
 \cite{Gibbons:2002bh} is based on the positive mass theorem. We speculate that the new argument using the divergence equations might be useful in order to obtain (\ref{yHbound}).

\subsection{Penrose inequality}

In the previous subsection, we have derived the Penrose-type inequality (\ref{Penroseineq_higher})
by evaluating the surface integral arising from the divergence equation (\ref{higherD_divJ}).
The appearance of the Penrose-type inequality rather than the reversed inequality is interesting and this might be helpful for the construction of the suitable flow in higher dimensions. 
As far as the Penrose inequality is concerned, we can derive it in several different fashions.

Setting $g_2(\rho)=0$ in the separable ansatz (\ref{ansatz}) and repeating the identical procedure, 
one can derive 
the following inequality
\begin{align}
\label{}
D^2 G-\frac 1V D^i VD_i G = & \frac{n-3}{2(n-2)}\left(\frac{\rho_0}{\rho}\right)^{\frac{n-3}{n-2}} 
\left|2\rho^2D_{[i}VH_{j]k}-\frac 2{n-2}H_{[i}g_{j]k} \right|^2\ge 0 \,,
\end{align}
where we have defined 
$G(V,\rho)\equiv V^2+(\rho_0/\rho)^{\frac{n-3}{n-2}}-1$. 
This is the equation for which the maximum principle can be applied~\cite{elliptic}, so that 
$G$ does not admit a maximum in the interior of $\Sigma$. 
Since $G=0$ both at infinity and at horizon, we have $G\le 0$
throughout $\Sigma$. 
Since $G $ is expanded at infinity as 
\begin{align}
\label{}
G=\frac{1}{r^{n-3}}(-2m+[(n-3)m\rho_0]^{\frac{n-3}{n-2}}) +O(1/r^{n-2}) \,,   
\end{align}
we conclude $(n-3)m \rho_0 \le (2m)^{\frac{n-2}{n-3}} $. 
Multiplying $\Omega_{n-2}$ and using Smarr's formula (\ref{Smarr_higher}), 
we readily obtain the Penrose inequality (\ref{Penroseineq_higher}).

\subsection{Modification of the proof based on the positive mass theorem}

As discussed in previous subsections, the quantity $\int_B {}^{(n-2)}R \D S$ is a primary 
obstruction of the present scheme in higher dimensions. 
Nevertheless, our formulation developed here is of use also for the uniqueness 
proof based upon the positive mass theorem \cite{bunting,Gibbons:2002bh}.  

To this aim, let us first recall the uniqueness argument by \cite{bunting,Gibbons:2002bh}, where the 
positive mass theorem has been ingeniously exploited to prove the conformal flatness of the 
constant timeslice. We first illustrate that the conformal transformation is imperative here. 
In terms of the isotropic coordinates,  
the higher-dimensional Schwarzschild metric can be written as 
\begin{align}
 \D s^2=& - \left[\frac{1-(\bar r_0/\bar r )^{n-3}}{1+(\bar r_0/\bar r)^{n-3}}\right]^2 \D t^2 +
\left[1+\left(\frac{\bar r_0}{\bar r }\right)^{n-3}\right]^{4/(n-3)}(\D \bar r^2+\bar r^2 \D \Omega_{n-2}^2) \,,
\label{Sch_isotropic}
\end{align}
where $\bar r_0 =(m/2)^{1/(n-3)}$.  By setting 
$V=\pm [1-(\bar r_0/\bar r )^{n-3}]/[1+(\bar r_0/\bar r)^{n-3}]$, the metric (\ref{Sch_isotropic})
is written as 
\begin{align}
\label{}
 \D s^2=-V^2 \D t^2+ \left(\frac{2}{1\pm V}\right)^{4/(n-3)}(\D \bar r^2+\bar r^2 \D \Omega_{n-2}^2) \,. 
\end{align}
This form of the metric manifests the conformal flatness of the constant timeslice. 
Bearing this form of the metric in mind, the authors in \cite{bunting,Gibbons:2002bh}  
considered two sort of the conformal transformations to $(\Sigma, g_{ij})$ as 
\begin{align}
\hat g_{ij}^{\pm}=\Omega_\pm ^2g_{ij}\,, \qquad
\Omega_\pm =\left(\frac{1\pm V}{2}\right)^{2/(n-3)} \,.
\label{BH_static_unique_CT}
\end{align}
One can easily check that  each manifold ($\Sigma_\pm, \hat g_{ij}^\pm$) is 
asymptotically Euclidean with the vanishing ADM mass and the vanishing scalar curvature. 
Glue these manifolds at $V=0$ and 
consider the complete Riemannian manifold $\bar \Sigma=\Sigma_+\cup\Sigma_-\cup\{\infty\} $. 
By positive mass theorem \cite{Schon:1979rg,SchonYau,Witten:1981mf}, $\bar \Sigma$ is flat.

The next step in~\cite{Gibbons:2002bh} is to embed the event horizon into the Euclid space $\mathbb E^{n-1}$.
Considering the local foliation of 
$\Sigma_+$ by $\ma S'_{v}=\{v\equiv 2/(1+V)={\rm constant}\}$ slice, 
the event horizon is located at $v=2$. It is easy to see that this surface $\ma S'_{v=2}$ is totally umbilic, viz, 
its second fundamental form is proportional to the first fundamental form with a constant coefficient. 
By the Gauss curvature decomposition, this kind of surface is maximally symmetric and of positive curvature, i.e, the induced metric on $\ma S'_{v=2}$ is a round sphere. Thus, the event horizon appears spherical when embedded in $\mathbb E^{n-1}$. 
Finally, one can conclude the spherical symmetry outside of the horizon
by noting that $v=2/(1+V)$ obeys a Laplace equation on $\mathbb E^{n-1}$, whose Dirichlet boundary value problem is unique.

Let us point out that the whole procedure in the previous paragraph can be by-passed. Once again, our tensorial quantity $H_{ij}$ provides more direct information on the underlying geometry.  An important fact here is that 
the Ricci tensor $\hat R_{ij}^\pm$ for the conformally transformed metric (\ref{BH_static_unique_CT}) must also vanish, 
when ($\Sigma_\pm, \hat g_{ij}^\pm$) is shown to be flat by the positive mass theorem. 
A simple calculation reveals 
\begin{align}
\label{RijHij}
\hat R_{ij}^\pm =\frac{1\mp V}{V(1\pm V)}H_{ij} \,.
\end{align}
It therefore follows that $H_{ij}$ defined in (\ref{Hij_higher}) is nothing but the Ricci tensor for the conformally transformed metric, up to a scalar function.
From (\ref{Sch_Uni_Hmndechigh}), the condition $H_{ij}=0$ implies $\sigma_{ij}=\ma D_i \rho=0$. 
The spherical symmetry on and outside the horizon immediately follows from the Gauss curvature decomposition formula in the space of the vanishing Weyl tensor. This standpoint is more geometric than previous analysis based on the Dirichlet boundary value problem.

\section{Summary and final remarks}
\label{sec:summary}

We made an extensive study on the uniqueness theorems of static black holes in the context of 
divergence equations. Following the strategy laid out in \cite{israel,MRS,robinson}, we have 
generalized these arguments into various directions. In the four dimensional vacuum case, 
our formula (\ref{vac_divid}) contains three tunable parameters, which allow us to conclude the spherical symmetry without resorting to the integrated mass formula. Using our divergence formula, one can 
also show $\chi\ge 2$. This is the inequality that is stronger than ever explored and excludes explicitly the real projective space. 
Furthermore, our tensorial field $H_{ij}$ defined in (\ref{vac_Hij}) enjoys 
a geometrically clear meaning, i.e., it describes the obstruction for the spherical symmetry (\ref{vac_Hij_dec}), the 
obstruction for the existence of the dilatation conformal Killing vector (\ref{Dizeta}), as well as the obstruction for the conformal 
Ricci flatness (\ref{RijHij}).  We expect that the discussion laid out in this paper will be applied for the stationary case, e.g., 
for the divergence equation in \cite{Simon:1983nm}. 

Our formulation is also robust in the four dimensional Einstein-Maxwell theory. 
For instance, one can apply the maximum/minimum principle to divergence-type equations to 
conclude that the electrostatic potential is a function of the norm of the static Killing vector. 
We believe that our formulation is insensitive to matter fields, as long as the material equations of motion are of 
divergence type. This is indeed the case for a theory with a conformally coupled scalar field \cite{tomikawa2017}. 

As we have verified, 
this advantage is optimized  in Einstein-Maxwell-dilaton theory. We found another divergence equation  (\ref{EMD_dividPhiPsi}), 
according to which we obtain the functional relationships for the norm of the Killing vector, electrostatic potential and 
dilaton field. This represents the effectiveness of the present scheme,  since 
it has been a long standing problem how to fix the value of the scalar field at the horizon. 
However, we do not know to what extent the coset representation comes into play for the existence of this type of divergence equation (\ref{EMD_dividPhiPsi}). It remains an intriguing issue to explore the case in which the scalar space is not symmetric nor homogeneous.

In higher dimensions, our divergence formula always encounters an intractable term $y_H$ given in (\ref{yamabe}). 
Although this limits the validity of the present strategy, it is still useful to obtain the Penrose-type inequality and 
for the modification of the uniqueness proof based upon the positive mass theorem. 
Interestingly, the bound (\ref{yHbound}) is the condition for the Penrose inequality for the time-symmetric Einstein-Maxwell initial data sets in higher dimensions, modulo some additional assumptions \cite{deLima:2014ysa}.  
This line of study is also worth exploring.

\acknowledgments
The authors are supported by Grant-in-Aid for Scientific Research from Ministry of Education, 
Science, Sports and Culture of Japan (17H01091). T.S. is supported by Grant-in-Aid for Scientific 
Research from Ministry of Education, Science, Sports and Culture of Japan (16K05344). M.N. is supported by 
Grant-in-Aid for Scientific Research (B) from JSPS (16H03979). 
K.I. is supported by Grant-in-Aid for Scientific Research from Ministry of Education, Science, Sports and 
Culture of Japan (17K14281). S.Y is supported by Grant-in-Aid for Scientific Research from Ministry of Education, Science, Sports and 
Culture of Japan (17H01091)

%======================================%
%<<<<<<<<<<<< REFERENCES  >>>>>>>>>>>>>%
%======================================%


\begin{thebibliography}{99}


\bibitem{israel} 
  W.~Israel,
  %``Event horizons in static vacuum space-times,''
  Phys.\ Rev.\  {\bf 164}, 1776 (1967).
%  doi:10.1103/PhysRev.164.1776
  %%CITATION = doi:10.1103/PhysRev.164.1776;%%
  
  
\bibitem{MRS}
H.~M\"uller Zum Hagen, D.~C.~Robinson and H.~J.~Seifert, 
%Black holes in static vacuum space-times
Gen.\ Rel.\ Grav.\ {\bf 4}, 53 (1973). 

  
  
\bibitem{robinson}
D.~C.~Robinson, 
%``A simple proof of the generalization of Israel's theorem,"
Gen.\ Rel.\ Grav.\ {\bf 8}, 696 (1977). 



  
\bibitem{bunting} 
G. L. Bunting and A. K. M. Masood-ul-Alam, Gen. Rel. Grav. {\bf 19}, 147 (1987).

\bibitem{Schon:1979rg} 
  R.~Schoen and S.~T.~Yau,
  %``On the Proof of the positive mass conjecture in general relativity,''
  Commun.\ Math.\ Phys.\  {\bf 65}, 45 (1979).
  %doi:10.1007/BF01940959
  %%CITATION = doi:10.1007/BF01940959;%%
  
  \bibitem{SchonYau} 
  R.~Schoen and S.~T.~Yau,
arXiv:1704.05490 [math.DG]. 
%``Positive Scalar Curvature and Minimal Hypersurface Singularities,''
  
  \bibitem{Witten:1981mf} 
  E.~Witten,
  %``A Simple Proof of the Positive Energy Theorem,''
  Commun.\ Math.\ Phys.\  {\bf 80}, 381 (1981).
  %doi:10.1007/BF01208277
  %%CITATION = doi:10.1007/BF01208277;%%
  
\bibitem{hwang}
S.~Hwang, 
Geometriae Dedicata {\bf 71}, 5 (1998).
%``A Rigidity Theorem for Ricci Flat Metrics,''



\bibitem{Gibbons:2002bh} 
  G.~W.~Gibbons, D.~Ida and T.~Shiromizu,
  %``Uniqueness and nonuniqueness of static vacuum black holes in higher dimensions,''
  Prog.\ Theor.\ Phys.\ Suppl.\  {\bf 148}, 284 (2003)
  %doi:10.1143/PTPS.148.284
  %[gr-qc/0203004].
  %%CITATION = doi:10.1143/PTPS.148.284;%%
  
\bibitem{Gibbons:2002av} 
  G.~W.~Gibbons, D.~Ida and T.~Shiromizu,
  %``Uniqueness and nonuniqueness of static black holes in higher dimensions,''
  Phys.\ Rev.\ Lett.\  {\bf 89}, 041101 (2002)
  %doi:10.1103/PhysRevLett.89.041101
  %[hep-th/0206049].
  %%CITATION = doi:10.1103/PhysRevLett.89.041101;%%


\bibitem{Gibbons:2002ju} 
  G.~W.~Gibbons, D.~Ida and T.~Shiromizu,
  %``Uniqueness of (dilatonic) charged black holes and black p-branes in higher dimensions,''
  Phys.\ Rev.\ D {\bf 66}, 044010 (2002)
  %doi:10.1103/PhysRevD.66.044010
  %[hep-th/0206136].
  
\bibitem{Rogatko:2003kj}
  M.~Rogatko,
  %``Uniqueness theorem of static degenerate and nondegenerate charged black holes in higher dimensions,''
  Phys.\ Rev.\ D {\bf 67} (2003) 084025
  %doi:10.1103/PhysRevD.67.084025
  %[hep-th/0302091].
  %%CITATION = doi:10.1103/PhysRevD.67.084025;%%
  %83 citations counted in INSPIRE as of 25 Apr 2018

\bibitem{Kunduri:2017htl}
  H.~K.~Kunduri and J.~Lucietti,
  %``No static bubbling spacetimes in higher dimensional Einstein-Maxwell theory,''
  Class.\ Quant.\ Grav.\  {\bf 35} (2018) no.5,  054003
%  doi:10.1088/1361-6382/aaa744
  %[arXiv:1712.02668 [gr-qc]].
  %%CITATION = doi:10.1088/1361-6382/aaa744;%%
  %1 citations counted in INSPIRE as of 25 Apr 2018
 
\bibitem{penrose1973} 
  R.~Penrose,
  %``Naked singularities,''
  Annals N.\ Y.\ Acad.\ Sci.\  {\bf 224}, 125 (1973).
%  doi:10.1111/j.1749-6632.1973.tb41447.x
  %%CITATION = doi:10.1111/j.1749-6632.1973.tb41447.x;%%





\bibitem{wald1977} 
P. S. Jang and R. M. Wald, J. Math. Phys. {\bf 18}, 41 (1977). 

\bibitem{imcf} 
G. Huisken and T. Ilmanen, J. Diff. Geom. {\bf 59}, 353 (2001). 


\bibitem{jang1979} 
  P.~S.~Jang,
  %``Note on cosmic censorship,''
  Phys.\ Rev.\ D {\bf 20}, 834 (1979).
%  doi:10.1103/PhysRevD.20.834
  %%CITATION = doi:10.1103/PhysRevD.20.834;%%

\bibitem{bray} 
  H. Bray, J. Diff. Geom. {\bf 59}, 177 (2001).


\bibitem{khuri2014} 
  M.~Khuri, G.~Weinstein and S.~Yamada,
  %``Proof of the Riemannian Penrose Inequality with Charge for Multiple Black Holes,''
  J. Diff. Geom. {\bf 106}, 451 (2017).
 % arXiv:1409.3271 [gr-qc].
  %%CITATION = ARXIV:1409.3271;%%

\bibitem{hawking1968} 
  S.~Hawking,
  %``Gravitational radiation in an expanding universe,''
  J.\ Math.\ Phys.\  {\bf 9}, 598 (1968).
%  doi:10.1063/1.1664615
  %%CITATION = doi:10.1063/1.1664615;%%

\bibitem{geroch1973} 
R. Geroch, Ann. N.Y. Acad. Sci. {\bf 224}, 108 (1973). 




  
  

 \bibitem{Mizuno:2009fj} 
  R.~Mizuno, S.~Ohashi and T.~Shiromizu,
  %``Static black hole uniqueness and Penrose inequality,''
  Phys.\ Rev.\ D {\bf 81}, 044030 (2010)
  %doi:10.1103/PhysRevD.81.044030
  %[arXiv:0911.5560 [gr-qc]].
  %%CITATION = doi:10.1103/PhysRevD.81.044030;%%
  
  %%%%%%%%%%%%%%%%%%%%%%%%%%%%%%%%%%%%
%  \bibitem{Lindblom}
%  L.~Lindblom and A.~K.~M.~Masood-ul-Alam, 
%   Comm.\ Math.\ Phys.  
%{\bf 162}, 123-145 (1994).
%On the spherical symmetry of static stellar models






  
%%%%%%%%%%%%%  vacuum %%%%%%%%%%%%%%%%%%
  \bibitem{Arnowitt:1962hi} 
  R.~L.~Arnowitt, S.~Deser and C.~W.~Misner,
  %``The Dynamics of general relativity,''
  Gen.\ Rel.\ Grav.\  {\bf 40}, 1997 (2008)
  %doi:10.1007/s10714-008-0661-1
  %[gr-qc/0405109].
  %%CITATION = doi:10.1007/s10714-008-0661-1;%%
  
  
  
 \bibitem{Hawking:1971vc} 
  S.~W.~Hawking,
  %``Black holes in general relativity,''
  Commun.\ Math.\ Phys.\  {\bf 25}, 152 (1972).
  %doi:10.1007/BF01877517
  %%CITATION = doi:10.1007/BF01877517;%%
  
    \bibitem{Chrusciel:1994tr} 
  P.~T.~Chrusciel and R.~M.~Wald,
  %``On the topology of stationary black holes,''
  Class.\ Quant.\ Grav.\  {\bf 11}, L147 (1994)
  %doi:10.1088/0264-9381/11/12/001
  %[gr-qc/9410004].
  

    
  
  \bibitem{Smarr:1972kt} 
  L.~Smarr,
  %``Mass formula for Kerr black holes,''
  Phys.\ Rev.\ Lett.\  {\bf 30}, 71 (1973)
  Erratum: [Phys.\ Rev.\ Lett.\  {\bf 30}, 521 (1973)].
  %doi:10.1103/PhysRevLett.30.71
  
    \bibitem{Lindblom}
  L.~Lindblom, 
%``Static uniform$B!>(Bdensity stars must be spherical in general relativity,''
J. Math. Phys. {\bf 29}, 436 (1988)
%doi:10.1063/1.528033




  
  

  
  
  %%%%%%%%%%%  Charged %%%%%%%%%%%
  \bibitem{Michalski} 
    H.~Michalski and J.~Wainwright, 
  %``Killing vector fields and the Einstein-Maxwell field equations in general relativity,''
  Gen.\ Rel.\ Grav.\  {\bf 6}, 289 (1975).

  \bibitem{Ray}
  J.~R.~Ray and E.~L.~Thompson, 
  J.\ Math.\ Phys.\ {\bf 16}, 345 (1975). 
%Spacetime symmetries and the complexion of the electromagnetic field

\bibitem{Tod:2006mp} 
  P.~Tod,
  %``Conditions for nonexistence of static or stationary, Einstein-Maxwell, non-inheriting black-holes,''
  Gen.\ Rel.\ Grav.\  {\bf 39}, 111 (2007)
  doi:10.1007/s10714-006-0363-5
  [gr-qc/0611035].
  %%CITATION = doi:10.1007/s10714-006-0363-5;%%
  
  
      \bibitem{Galloway1995}
  G.~ J.~Galloway, 
  Class.\ Quant.\ Grav. {\bf 12}, L99 (1995). 
  %On the topology of the domain of outer communication
  
  
    \bibitem{Gibbons:1982fy} 
  G.~W.~Gibbons and C.~M.~Hull,
  %``A Bogomolny Bound for General Relativity and Solitons in N=2 Supergravity,''
  Phys.\ Lett.\  {\bf 109B}, 190 (1982).
  %doi:10.1016/0370-2693(82)90751-1
  %%CITATION = doi:10.1016/0370-2693(82)90751-1;%%
  
  \bibitem{Nozawa:2014zia} 
  M.~Nozawa and T.~Shiromizu,
  %``Positive mass theorem in extended supergravities,''
  Nucl.\ Phys.\ B {\bf 887}, 380 (2014)
  %doi:10.1016/j.nuclphysb.2014.09.002
  %[arXiv:1407.3355 [hep-th]].
  
  
  
  \bibitem{Israel:1967za} 
  W.~Israel,
  %``Event horizons in static electrovac space-times,''
  Commun.\ Math.\ Phys.\  {\bf 8}, 245 (1968).
  %doi:10.1007/BF01645859
  %%CITATION = doi:10.1007/BF01645859;%%
  
  \bibitem{MR}
H.~M\"uller Zum Hagen and D.~C.~Robinson, 
%Black holes in static electrovac space-times
Gen.\ Rel.\ Grav.\ {\bf 5}, 61 (1974). 

\bibitem{Simon1984}
W.~Simon, 
Gen.\ Rel.\ Grav.\ {\bf 17}, 761 (1984). 
%A Simple Proof of the Generalized Electrostatic Israel Theorem 

\bibitem{Ruback}
P.~Ruback, 
%A new uniqueness theorem for charged black holes
Class.\ Quant.\ Grav.\  {\bf 5}, L155 (1988). 


\bibitem{MuA1992}
K.~M.~Masood-ul-Alam, 
Class.\ Quant.\ Grav.\  {\bf 9}, L53 (1992). 
%Uniqueness proof of static charged black holes revisited

  

  \bibitem{elliptic}
D.~Gilbarg and N.~S.~Trudinger, {\it ``Elliptic Partial Differential Equations of Second Order,''}
Springer (1977). 
  
%%%%%%%%%%%%%  dilaton   %%%%%%%%%%%%%


  \bibitem{Gibbons:1993xt} 
  G.~W.~Gibbons, D.~Kastor, L.~A.~J.~London, P.~K.~Townsend and J.~H.~Traschen,
  %``Supersymmetric selfgravitating solitons,''
  Nucl.\ Phys.\ B {\bf 416}, 850 (1994)
  %doi:10.1016/0550-3213(94)90558-4
  %[hep-th/9310118].
  %%CITATION = doi:10.1016/0550-3213(94)90558-4;%%
  
  
  \bibitem{Nozawa:2010rf} 
  M.~Nozawa,
  %``On the Bogomol'nyi bound in Einstein-Maxwell-dilaton gravity,''
  Class.\ Quant.\ Grav.\  {\bf 28}, 175013 (2011)
  %doi:10.1088/0264-9381/28/17/175013
  %[arXiv:1011.0261 [hep-th]].
  %%CITATION = doi:10.1088/0264-9381/28/17/175013;%%


\bibitem{Gibbons:1987ps} 
  G.~W.~Gibbons and K.~i.~Maeda,
  %``Black Holes and Membranes in Higher Dimensional Theories with Dilaton Fields,''
  Nucl.\ Phys.\ B {\bf 298}, 741 (1988).
  %doi:10.1016/0550-3213(88)90006-5
  %%CITATION = doi:10.1016/0550-3213(88)90006-5;%%
  
  
\bibitem{MuA:dilaton}
A.~K.~M.~Masood-ul-Alam, 
Class.\ Quant.\ Grav. {\bf 10}, 2649 (1993).
%Uniqueness of a static charged dilaton black hole


%\cite{Mars:2001pz}
\bibitem{Mars:2001pz} 
  M.~Mars and W.~Simon,
  %``On uniqueness of static Einstein-Maxwell dilaton black holes,''
  Adv.\ Theor.\ Math.\ Phys.\  {\bf 6}, 279 (2003)
  %doi:10.4310/ATMP.2002.v6.n2.a3
  %[gr-qc/0105023].
  %%CITATION = doi:10.4310/ATMP.2002.v6.n2.a3;%%
  
  
%%%%%%%%%%%%%%%%  Summary   %%%%%%%%%%%%%%%%%
  
  \bibitem{tomikawa2017} 
  Y.~Tomikawa, T.~Shiromizu and K.~Izumi,
  %``On uniqueness of static black hole with conformal scalar hair,''
  Prog.\ Theor.\ Exp.\ Phys. {\bf 2017}, 033E03 (2017).
  %arXiv:1612.01228 [gr-qc].
  %%CITATION = ARXIV:1612.01228;%%
  
  
  \bibitem{Simon:1983nm} 
  W.~Simon,
  Gen.\ Rel.\ Grav.\ {\bf 16}, 465 (1983).
  %``Characterizations Of The Kerr Metric,''
  
  \bibitem{deLima:2014ysa} 
  L.~Lopes de Lima, F.~Girao, W.~Lozorio and J.~Silva,
  %``Penrose inequalities and a positive mass theorem for charged black holes in higher dimensions,''
  Class.\ Quant.\ Grav.\  {\bf 33}, no. 3, 035008 (2016)
  %doi:10.1088/0264-9381/33/3/035008
  %[arXiv:1401.0945 [math.DG]].
  %%CITATION = doi:10.1088/0264-9381/33/3/035008;%%
  
  
\end{thebibliography}
\end{document}